\documentclass[a4paper,11pt]{article}
\usepackage{amsmath}
\usepackage{amssymb}
\usepackage{amsthm}
\usepackage{thmtools}
\usepackage[round]{natbib}
\usepackage{color}
\usepackage{graphicx}
\usepackage{comment}
\usepackage{hyperref}
\hypersetup{
    unicode=false,          
    pdftoolbar=true,        
    pdfmenubar=true,        
    pdffitwindow=false,     
    pdfstartview={FitH},    
    colorlinks=true,       
    linkcolor=black,          
    citecolor=blue,        
    filecolor=magenta,      
    urlcolor=black           
}
\usepackage{adjustbox}
\usepackage[section]{placeins}
\usepackage{fullpage}
\usepackage{caption}
\usepackage{subcaption}
\usepackage[ruled]{algorithm2e}
\usepackage{float}

\usepackage{tikz}
\usetikzlibrary{calc}

\newtheorem{theorem}{Theorem}
\newtheorem*{example}{Example}
\newtheorem{assumption}{Assumption}

\newcommand{\dred}[1]{#1}

\newcommand{\argmin}[1]{\underset{#1}{\operatorname{arg}\,\operatorname{min}}\;}
\newcommand{\argmax}[1]{\underset{#1}{\operatorname{arg}\,\operatorname{max}}\;}

\def\1{\mathbf{1}}

\newcommand{\RR}{\mathbb R}

\newcommand{\Var}{\text{var}}
\newcommand{\Ss}{S}
\newcommand{\Esp}{\text{E}}
\newcommand{\R}{{r}}

\newcommand{\Set}{\mathcal{F}}

\title{Optimal sequential sampling design for environmental extremes}
\author{Rapha\"el de Fondeville and Matthieu Wilhelm}

\begin{document}
\maketitle
\begin{abstract}
The Sihl river, located near the city of Zurich in Switzerland, is under continuous and tight surveillance as it flows directly under the city's main railway station.
To issue early warnings and conduct accurate risk quantification, a dense network of monitoring stations is necessary inside the river basin.
However, as of $2021$ only three automatic stations are operated in this region, naturally raising the question: how to extend this network for optimal monitoring of extreme rainfall events?

So far, existing methodologies for station network design have mostly focused on maximizing interpolation accuracy or minimizing the uncertainty of some model's parameters estimates.
In this work, we propose new principles inspired from extreme value theory for optimal monitoring of extreme events.
For stationary processes, we study the theoretical properties of the induced sampling design that yields non-trivial point patterns resulting from a compromise between a boundary effect and the maximization of inter-location distances.
For general applications, we propose a theoretically justified functional peak-over-threshold model and provide an algorithm for sequential station selection.
We then issue recommendations for possible extensions of the Sihl river monitoring network, by efficiently leveraging both station and radar measurements available in this region.
\end{abstract}

\section{Introduction}\label{sec: intro}
\begin{figure}
\begin{center}
\includegraphics[width=0.6\textwidth]{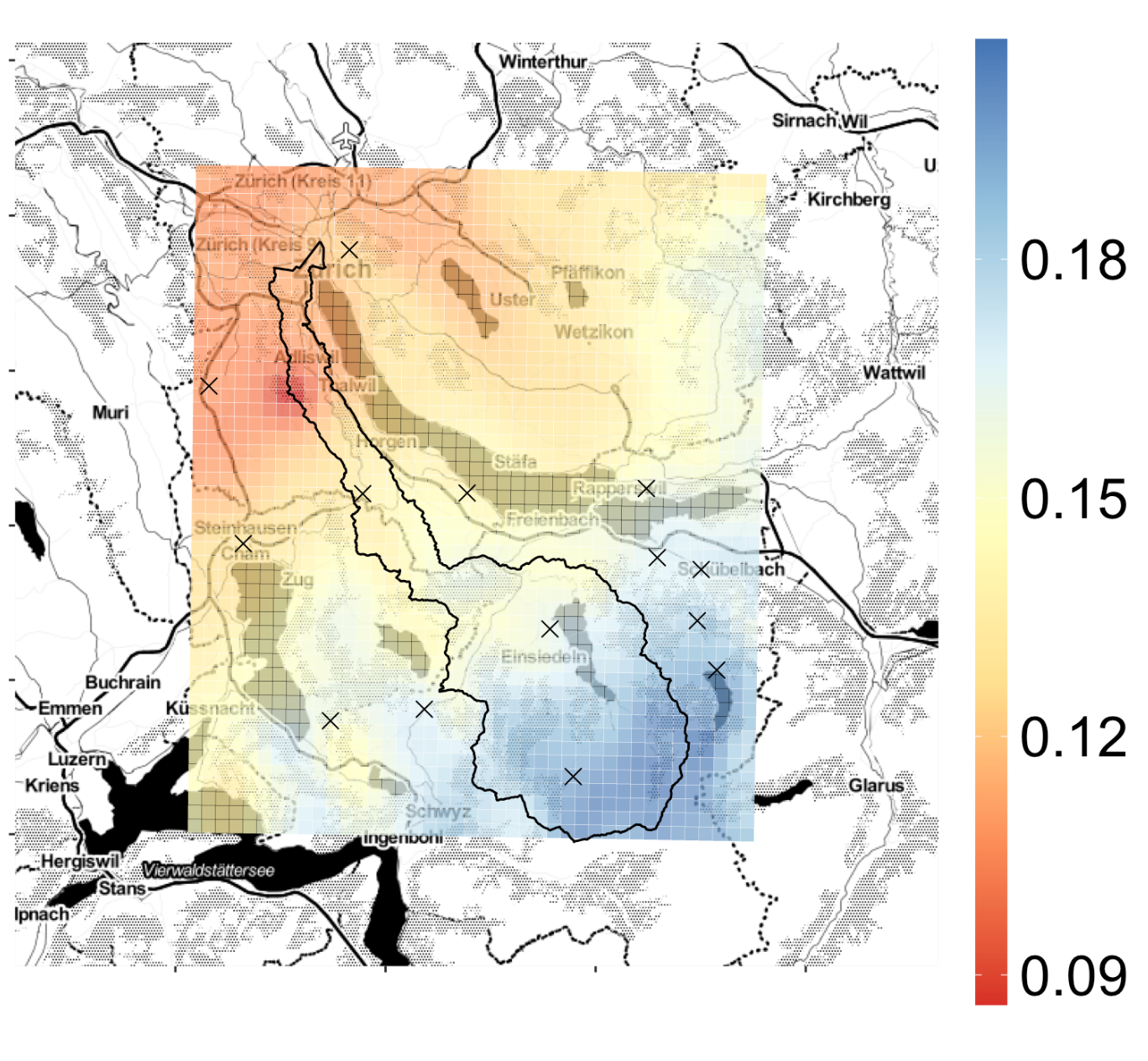}
\end{center}
\caption{Hourly mean rainfall (mm) from radar measurements over the period $2013$ to $2018$ around the Sihl river basin. The solid black line delimits the Sihl river basin while black crosses represent the locations of the $14$ weather stations installed by MeteoSwiss.}\label{fig: basin and stations}
\end{figure}

During summer $2005$, the city of Zurich was heavily flooded causing six deaths and inducing an estimated property damage of around $3$ billions Swiss francs  \citep{Bezzola2007}.
As illustrated by Figure~\ref{fig: basin and stations}, the lake side location of Zurich makes the city particularly prone to flood risk, especially that several rivers flow directly through its center.
One of these rivers, the Sihl, is particularly monitored as its waterbed is located directly under the city main railway station, which is partly built underground.
Thus Sihl extreme water heights are likely to cause hundreds of millions of francs of losses by damaging existing infrastructures and by preventing half a million commuters from travelling.

Following the $2005$ flood, which fortunately spared the Sihl, an overall assessment of flood risk for the city has been conducted, leading to a new local policy for disaster risk reduction.
As a consequence, civil engineering infrastructures to minimize long term risk exposure were constructed. 
These were combined to short term mitigation measures, namely a system of ``early warnings'', which relies on weather forecasts and delivers to populations and authorities messages of likely flooding in the coming hours.
In both cases, these protective measures require to monitor rainfall extremes as accurately as possible, to adequately calibrate the infrastructure in the first case, and minimize false alarms in the second. 

To this end, the main sources of rainfall measurements are weather stations, represented by crosses in Figure \ref{fig: basin and stations}. Of all sources, this kind of observations have usually the longest temporal coverage: for instance in Switzerland, oldest rainfall records dates from $1863$.
These measures are however sparse in space, only $14$ around the Sihl river with only $3$ of them located inside the basin, and may suffer from multiples sources of uncertainties such as weather conditions, e.g., wind or snow, right censoring or instruments changes.
As the only source of direct measurements, station data is in general considered as most accurate for risk quantification.

With the raise of new technologies, radar measurements are now much more frequently available: for instance ``CombiPrecip'' produced by MeteoSwiss \citep{Sideris2014a,Gabella2017,Panziera2018} provides spatially dense estimates of temporally accumulated rainfall at high spatial resolution since $2013$.
These observations are however indirect and result from processing radio waves acquisitions, which are usually subject to strong bias and distortion, especially in region with altitude variation.
There is usually a large discrepancy between measurements provided by radar products and by weather stations, and their relationship is today far from being understood.
For this reason, radar products are, as of today, almost exclusively used for short term mitigation and need to be combined to station measurements to correct potential biases in intensity.

Techniques for flood risk quantification, both long and short term, relies on catalogues of heavy rainfall events that are then fed to a hydrological numerical model which translates the event into water runoffs.
These catalogues are either historical, i.e., derived from available measurements, or produced by weather generators that can be either numerical  \citep{Cloke2009} or stochastic \citep[see, e.g.,][]{Furrer2008,DeFondeville2017}; in the latter extreme value theory offers a mathematically justified framework to model and extrapolate extreme water levels.
In both cases, the methodologies extrapolate stations measurements away from monitoring sites to estimate the water distribution over the region of interest.
The accuracy of such techniques is thus tightly linked to the number of monitoring stations present in the basin.
For the Sihl river, as of $2021$, only three stations provides automatic rainfall measurements, and because of the importance of the infrastructure at risk, it would be necessary to extend the network of station for optimal monitoring of extreme rain events.

In this work, we define and explore sampling designs for optimal monitoring of extremes events, with the goal to efficiently extend the existing network of stations in the Sihl river basin.
More precisely, we propose a model based on extreme value theory that leverages the strength of both station and radar measurements and issue recommendations for candidate locations of new stations.
Section \ref{sec: principles sampling design} reviews classical principles for sampling design over compact spaces, proposes a new paradigm for extreme monitoring and derives its properties for stationary random fields.
Section \ref{sec: evt model} presents functional peaks-over-threshold analysis by describing the asymptotic tail behaviour of stochastic processes, that is then combined with the previously introduced principle for sampling design to provide explicit model and algorithm to optimal monitoring of extremes.
Section \ref{sec: sihl application} details the Sihl case study and provides recommendation to extend the current network of measurement stations.
Finally, Section \ref{sec: discussion} concludes with the limitations of the current model and possible improvements for future research.

\section{Monitoring extremes of random fields on compact sets}\label{sec: principles sampling design}
\subsection{Related literature}
Stations networks are installed to fulfil one or several, possibly contradicting, monitoring purposes \citep{Chang2007}: their design can be thus formulated as an optimization problem, potentially under constraints such as budget or landscapes regulations, with a clear objective function to be minimized \citep[p. 502,][]{Zidek2019}.
Depending on choice of objective, data availability, and, field assumptions, network designs are classified as geometry, probability and model based.
All three categories rely on the minimal assumption that the process $X$ to be monitored is continuous over the region $S$ of interest, that we assume bounded.

The first approach, also known as space-filling designs, exploits only the geometrical properties of $S$, and as such, is useful when little to no information is available about the monitoring purpose object; see, e.g., \citet{Pronzato2012} for an extensive review of such designs.
Networks derived from this principle turned out to be performing well in settings where dependence is known and the network purpose is limited to spatial prediction \citep{Li2015}; they can perform quite poorly in other contexts.

Probability based designs, also known simply as `design-based' in survey sampling, consider the object of monitoring to be deterministic, but makes stations locations random by imposing a model on their distribution; see \citet{Tille2017} for a recent review of such techniques.
One of their appealing property is the intrinsic agnostic nature of the design: stations are selected with the help of a prescribed distribution and no modelling assumption is made on the field $X$.
For this reason, probability based designs are often considered as `free' from any kind of subjective bias induced by any prior knwoledge on $X$ or the geometry of $S$.
These techniques have been mostly developed for optimal estimation of the mean or the cumulative distribution function of the process.
While they can be used outside this scope, their performance is usually not ensured.

The third, and last, class of designs assumes that $X$ is a continuous stochastic process over $S$ whose distribution is given by a model imposed by the designer and tailored to reflect the specificities of the network purpose(s).
In this case, the set of stations induced by the sampling design is skewed by the incorporation of prior knowledge, making its performance determined by the model's validity.
In this context, an attractive objective function is one that selects locations such that model's parameters estimates, e.g., covariance functions parameters \citep{Muller1999}, have minimal uncertainty.
However, in general, there is a trade-off between optimal parameters estimation and minimal prediction error at unmeasured locations.
\citet{Zidek2019} argues that entropy, which aims at optimally reducing the uncertainty of predictions at unmeasured locations, achieves a compromise between both previous objectives. 

In absence of collected data, a setting that is known as ``optimal design of experiments'' in statistics, model estimation is likely to not be possible, and we need to make potentially strong hypotheses on the distribution of $X$, the most common being strict stationarity.
\citet{Pukelsheim06} provides an exhaustive review of this branch of statistics, that \citet{Zidek2019} refers to as \textit{de novo} design.
Alternatively, when data is available, a more general model, whose flexibility depends on the quantity and quality of available data, can be fitted to compute optimal sets of measurement stations.

Model based designs leveraging existing data are tightly linked to the literature on design of computer experiment \citep{Santner2018}, which
mostly relies on Gaussian processes.
Theoretical properties of Gaussian designs have been studied for multiple purposes such as global optimization \citep{Schonlau98}, environmental risk assessment \citep{Arnaud10, Chevalier14techno, Azzimonti16, Azzimonti19} and uncertainty quantification for level sets \citep{Azzimonti19b}. 
In particular, \citet{Bect2012} propose network designs to detect probabilities of failure, i.e., finding regions, called excursion sets, where the probability of the Gaussian process $X$ to exceeds a threshold $u$ is greater than some quantile of reference \citep{Adler07};
subsequent research in the same area includes \citet{Chevalier14techno} and \citet{Azzimonti16, Azzimonti19}.
While Gaussian processes are attractive for versatility and convenience, assuming normal distributions is likely to strongly under-estimate rare extreme events with potentially dire consequences for purposes such as risk quantification of potentially heavy-tailed processes.

The literature on network design for monitoring of extreme events is, to this date, rather limited: \citet{Chang2007} discuss the challenges arising in the design of networks for extremes monitoring.
For this specific purpose, they give evidences against classical designs and propose a Bayesian hierarchical model with a Gaussian copula with good empirical performance, but do not ``appeal to an axiomatic theory as in the classical theory of extremes'', which is the goal of the current work.
More recently, \citet{Hainy2016} propose a model for optimal monitoring of yearly maxima using extreme value theory that they use to rank existing stations within a network as function of their impact on the estimation accuracy.
In this work, we propose to monitor single extreme events, defined as specific types of exceedances, instead of maxima and propose a principled approach to the general problem of sampling designs for optimal monitoring of extremes.

\subsection{Principle of sampling designs for extremes}

When studying extremes, the definition of an objective function for the optimization of the sampling process must be carefully considered. Indeed, variance reduction under an unbiasedness constraint, which is classically used for sampling design, is not a good criterion: first it sets the focus on the accurate estimation of the process mean, and second, variance might even not exist if the process is sufficiently heavy-tailed.
For this reason, we propose an alternative criterion for extremal sampling design based on a risk measure.

For univariate quantities, risk can be characterized by the distribution of exceedances over a high threshold, whose value can be chosen, for instance, as the minimum over which damages occur.
For random fields, the notion of exceedance, and so of risk, is however not unique.
In this case, we rely on the notion of $\R$-exceedance \citep{Dombry2013} for which a stochastic process $X$ over a compact domain $S$ is summarized by a univariate summary statistics $\R(X)$ computed with the help of a risk functional $\R$.
In this case, an extreme event is defined as an exceedance of the functional $\R(X)$ over the threshold $u$.
The functional $\R$ characterizes the risk under study, for instance, in case of rainfall, it may distinguish different underlying physical processes, e.g., cyclonic or convective rain.
Indeed, both types of rainfall event are driven by different physical laws, producing rain fields with different spatio-temporal structures.
The role of risk functional in this case is thus to disentangle both type of extreme events, see \citet{DeFondeville2017} for a detailed illustration.

Popular choices of risk functionals are for instance $\R(X) = \sup_{s \in \Ss} X(s)$, where events with at least one location above a threshold are considered as extremes or $\R(X) = \int_0^T \int_\Ss X(s,t)dsdt$, which computes, when $X$ represents the rain field, the volume of water fallen over the region $\Ss$ over the time window $[0,T]$. In any case, it defines a measure of the severity of the event. 

Suppose now that we can evaluate the process $X$ only at a limited number of sites $L > 0$, determined for instance by the available budget to install measuring equipment. 
We wish to find an optimal set of $L$ locations, say $S_{\rm samp} \subset S$ such that the risk is best quantified, i.e., such that
\begin{equation}\label{eq: samp princ}
S_{\rm samp} = \underset{\{s_1,\dots,s_L\} \subset S}{\rm argmin} \left|\Pr\left\{ \R(X) \geq u\right\} - \Pr\left\{ \R_{\{s_1,\dots,s_L\}}(X) \geq u| \R(X) \geq u\right\}\right|,
\end{equation}
where $\R_{\{s_1,\dots,s_L\}}(X)$ is a consistent estimator of $\R(X)$, i.e., a quantity derived from the vector $\{X(s_1), \dots, X(s_L)\}$ and such that $\R_{\{s_1,\dots,s_L\}}(X) \rightarrow \R(X)$ as $L \rightarrow \infty$, in some sense.
Simple examples of discretization of $r(X) = \sup_{s\in \Ss} X(s)$  and $r(X) = \int_{\Ss} X(s) \ ds$ are $\max_{s \in S_{\rm samp}} X(s)$ and $L^{-1}\sum_{i\in \Ss_{\rm samp}} X(s_i)$ respectively, and the convergence (in probability) occurs for instance if $S_{\rm samp}\stackrel{\text{i.i.d}}{\sim} \mathcal{U}(\Ss)$, where $\mathcal{U}$ denotes the uniform distribution on $\Ss$.
Thus, sampling design minimizing \eqref{eq: samp princ} will provide a set of locations for which the probability of the risk $\R(X)$ to exceeds threshold $u \in \RR$ is best approximated using the process sampled at locations in $S_{\rm samp}$.

A sample $S_{\rm samp}$ as defined by \eqref{eq: samp princ} is meaningful for both long and short term risk mitigation: the construction of civil engineering infrastructure necessitates an adequate quantification of the risk and early warnings are usually defined by hazards levels, which are derived form probability of exceeding certain levels of intensity.
In the latter case, false negative are likely to have dire consequences, as population might not be informed of an imminent natural disaster, and thus it is critical that the network of station minimizes \eqref{eq: samp princ} as much as possible.
For risk estimation, the analysis relies in general on a model estimated using all observation for which there is a significant risk, i.e., using the previous definition, such that $\R(X) > u$.
Thus accurate risk estimation is tightly linked by our capability to accurately estimate $\Pr\{\R(X) \geq u\}$ from its discretization.

Finally, we could also have considered alternative criteria such as minimizing the absolute difference between $\Esp\{\R(X)\}$ and the expectation of its estimator for finite $L$. This choices would however be less general as it requires to suppose the existence and finiteness of such quantities, which is non-trivial to prove for most of existing random fields.

\subsection{Properties of sampling designs for extremes of stationary processes}\label{sec: properties extr sampling}
We now study the properties of the sampling design over a compact domain $\Ss \subset \mathbb{R}^d$ induced by the equation \eqref{eq: samp princ} for stationary processes $X$.
Although rarely satisfied in practice, stationarity is a convenient working hypothesis when little to no a priori information about the tail distribution of the data is available or when trying to understand the theoretical properties of stochastic processes.
We thus suppose that $X$ is strictly stationary, i.e., for any locations $s_1, \dots, s_L \in \Ss$, direction \dred{$h$ in the $d-1$ unit sphere $\mathbb{S}^{d-1}$}, and scalar $t >0$, we assume that
\begin{equation*}\label{eq:stationarity}
\Pr\{ X(s_1) \leq x_1, \dots, X(s_L) \leq x_L\} = \Pr\{ X(s_1 + th) \leq x_1, \dots, X(s_L + th) \leq x_L\}, \quad  x \in \RR^L.
\end{equation*}
From the point of view of extremes, stationarity implies that for any threshold $(u_1,u_2) \in \RR^2$,
\begin{equation}\label{eq: hyp de novo}
\pi(S_1,S_2) = \Pr\{\sup_{s \in S_1} X(s) \geq u_1, \sup_{s \in S_2} X(s) \geq u_2\}, \quad S_1,S_2 \subset \RR^d,
\end{equation}
is invariant by translation of $S_1$ and $S_2$.
The function $\pi$ can be used as a measure of spatial dependence and its limit for increasingly large threshold when divided by $ \Pr\{\sup_{s \in S_1} X(s) \geq u\}$ is commonly used to quantify extremal dependence \citep{Trust2017,Engelke}. 

To describe the properties of the sampling design induced by \eqref{eq: samp princ}, we will assume that $\pi$ is a decreasing function function of the distance between $S_1$ and $S_2$, i.e., the probability that the process $X$ exceeds a potentially large threshold simultaneously on two regions, decreases as a function of the distance separating them.
This hypothesis is natural for environmental applications and translates, for the analysis of extremes, the First Law of Geography \citep{Tobler1970}, i.e., `everything is related to everything else, but near things are more related than distant things'.
In classical spatial statistics, an equivalent hypothesis would simply assume that the covariance function is decreasing with the distance.

To avoid any contradiction with Tobler's law, the probabilistic nature of $\pi$ requires further assumptions to discard the possibility to find conditions for which there exists a vector $h_1 \in \mathbb{S}^{d-1}$ and $t > 1$ such that $\pi(S_1,S_2 + th_1) > \pi(S_1, S_2 + h_1)$.
To do so, we introduce the notion of hitting scenario:
for a threshold $u$, we call a hitting scenario a function $H$ which for any pair $(S_1,S_2) \subset \Ss \times \Ss$ associates a partition $H(S_1,S_2) = \{H_k\}_{k = 1}^K$ of the set $\{x \in C(\RR^d) : \sup_{s \in S_1} X(s) \geq u, \sup_{s \in S_2} X(s) \geq u\}$.
For instance, a simple hitting scenario distinguishes between random paths over $\Ss$ for which $ \sup_{s \in S_1} X(s) > \sup_{s \in S_2} X(s)$.
In other words, these scenarios allow to characterize more precisely the context in which joint exceedances take place, similarly to the work of \cite{Wang2013} for maxima: both notions share the underlying idea to distinguish between the different paths yielding to a same observed event.

\begin{assumption}
\label{assump:pi_decreasing}
For any convex $\Ss \subset \RR^d$, vector $h \in \mathbb{S}^{d-1}$ and hitting scenario $H$, 
$$
\pi_H(\Ss, \Ss + th) =  \Pr\{\sup_{s \in \Ss} X(s) \geq u, \sup_{s \in \Ss + th} X(s) \geq u, X \in H_k\}, \quad k = 1,\dots, K$$
is a decreasing function of $t > 0$.
\end{assumption}


{Intuitively, Assumption \ref{assump:pi_decreasing} formally translates the first law of geography for the dependence measure \eqref{eq: hyp de novo}.
Omitting hitting scenarios would not ensure that, when analyzing the different paths yielding joint exceedances, it is not possible to find a configuration under which there exists $t > 1$ and $h\in \mathbb{S}^{d-1}$ such that $\pi_H(\Ss, \Ss + th) > \pi_H(\Ss, \Ss + h)$, for which Tobler's law is obviously violated. 

\begin{example}
Let $X$ be a continuous stationary Gaussian process with decreasing covariance function $\sigma$ and mean 0.
Consider $S_1 = \{s_1\} \subset \Ss$ and $S_2 = \{s_2\} \subset \Ss$ and the hitting scenario
 $$H:(s_1,s_2) \rightarrow \{x \in C(S): (-1)^i \times\{ x(s_1) > x(s_2)\}\}_{i = 1,2},$$
then
$$
\pi_H\left[\{s_1\},\{s_2\}\right] =  \Phi_{0,\Sigma}(-u,-u) / 2,
$$
where $\Phi_{0,\Sigma}$ is the distribution function of a bivariate normal random variable with zero mean and covariance $\Sigma = \{\sigma(s_i - s_j)\}_{i,j= 1,2}$, and thus $\pi_H$ is a decreasing function of the distance between $S_1$ and $S_2$.
A similar formula can be derived for generalized $\R$-Pareto process; see Section \ref{sec: gen Pareto theory}.
\end{example}}

In some cases, the risk functional and its discretization can be ordered:
for instance, the supremum over $\Ss$ is always greater than its discretization, and \eqref{eq: samp princ} simplifies to
\begin{align}\label{eq: de nove princ simp}
\Ss_{\rm samp} & = \underset{ \{s_1,\dots,s_L\}}{\rm argmax} \Pr\left.\left\{ \max_{s \in \{s_1,\dots,s_L\}}X(s) \geq u\right| \sup_{s \in \Ss} X(s) \geq u\right\}.
\end{align}
Similar simplifications, with possibly reverse relations, also holds for the infimum.
For the analysis of extremes, the supremum functional is a common choice that we choose to focus on: for locations $s_1, \dots, s_{L}\in \Ss$, we observe that
\begin{align}
 &\Pr\left.\left\{ \max_{s \in \{s_1,\dots,s_L\}}X(s) \geq u\right| \sup_{s \in \Ss} X(s) \geq u\right\}\nonumber \\
 & =\sum_{i = 0}^{L-1}\Pr\left\{X(s_{i+1})\geq u \left|  \sup_{s \in \Ss} X(s) \geq u\right\}\right. -\sum_{i = 1}^{L-1} \Pr\left\{ X(s_{i+1}) \geq u, \max_{s \in \{s_1,\dots,s_{i}\}}X(s) \geq u \left| \sup_{s \in \Ss} X(s) \geq u \right\}\right.  \label{eq: decomposition spacefilling}\\
 & = \Pr\left\{X(s_1)\geq u \left|  \sup_{s \in \Ss} X(s) \geq u\right\}\right. + \sum_{i = 1}^{L - 1} \Pr\left\{ X(s_{i + 1}) \geq u, \max_{s \in \{s_1,\dots,s_{i}\}}X(s) \leq u \left| \sup_{s \in \Ss} X(s) \geq u \right\}\right., \label{eq: decomposition boundary}
\end{align}
allowing to elucidate the nature of optimal sampling designs induced by criteria \eqref{eq: de nove princ simp}.
Because of stationarity, marginal probabilities in \eqref{eq: decomposition spacefilling} do not influence the design choice, so, at first sight, solving \eqref{eq: de nove princ simp} under Assumption \ref{assump:pi_decreasing} seems equivalent to maximize inter-points distances, i.e., an optimal sampling design would, roughly speaking, cover $\Ss$ as extensively as possible.
However, condition $\sup_{s \in \Ss} X(s) \geq u$ induces a boundary effect, whose strength is function of the dependence, making terms in \eqref{eq: decomposition boundary} to be influenced not only by the distances between sampling locations but also by their position with respect to the boundary $\partial \Ss$ of $\Ss$.

\begin{figure}
    \centering
    \begin{tabular}{cc}
    {\footnotesize $\Pr \left.\left\{\underset{s \in \dred{[0,1]}}{\sup} X(s) \geq 1, \underset{s \in [h,h + 1]}{\sup} X(s) \geq 1 \right| \dred{\underset{s \in \dred{[0,1]}}{\sup} X(s) \geq 1}\right\}$} & {\footnotesize$\Pr\left.\left\{\underset{s \in S_h}{\sup} X(s) \geq 1, \underset{s \in S\setminus  S_h}{\sup} X(s) < 1 \right| \dred{\underset{s \in \Ss}{\sup\ } X(s) > 1} \right\}$}\\~\\
        \includegraphics[width=0.4\textwidth]{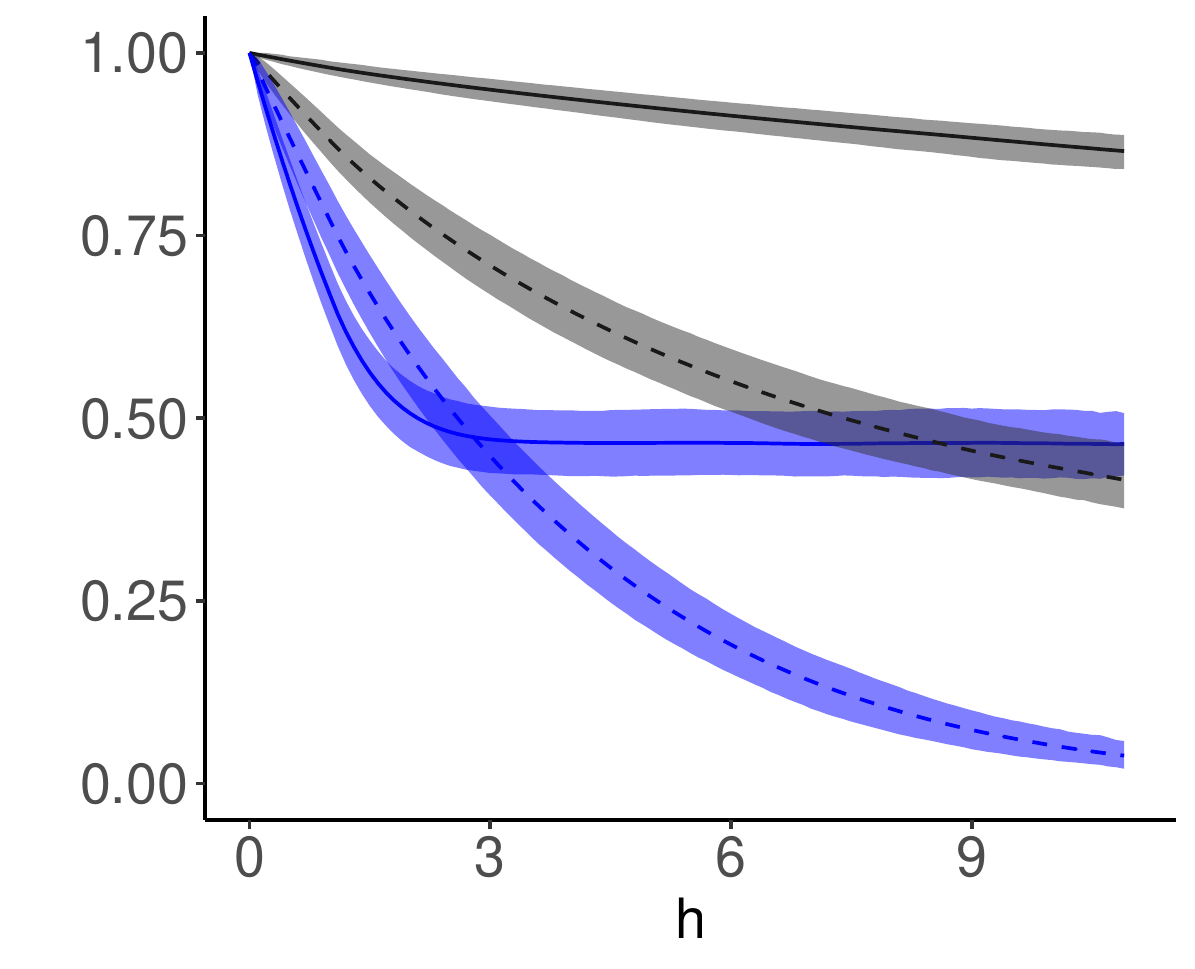} & \includegraphics[width=0.4\textwidth]{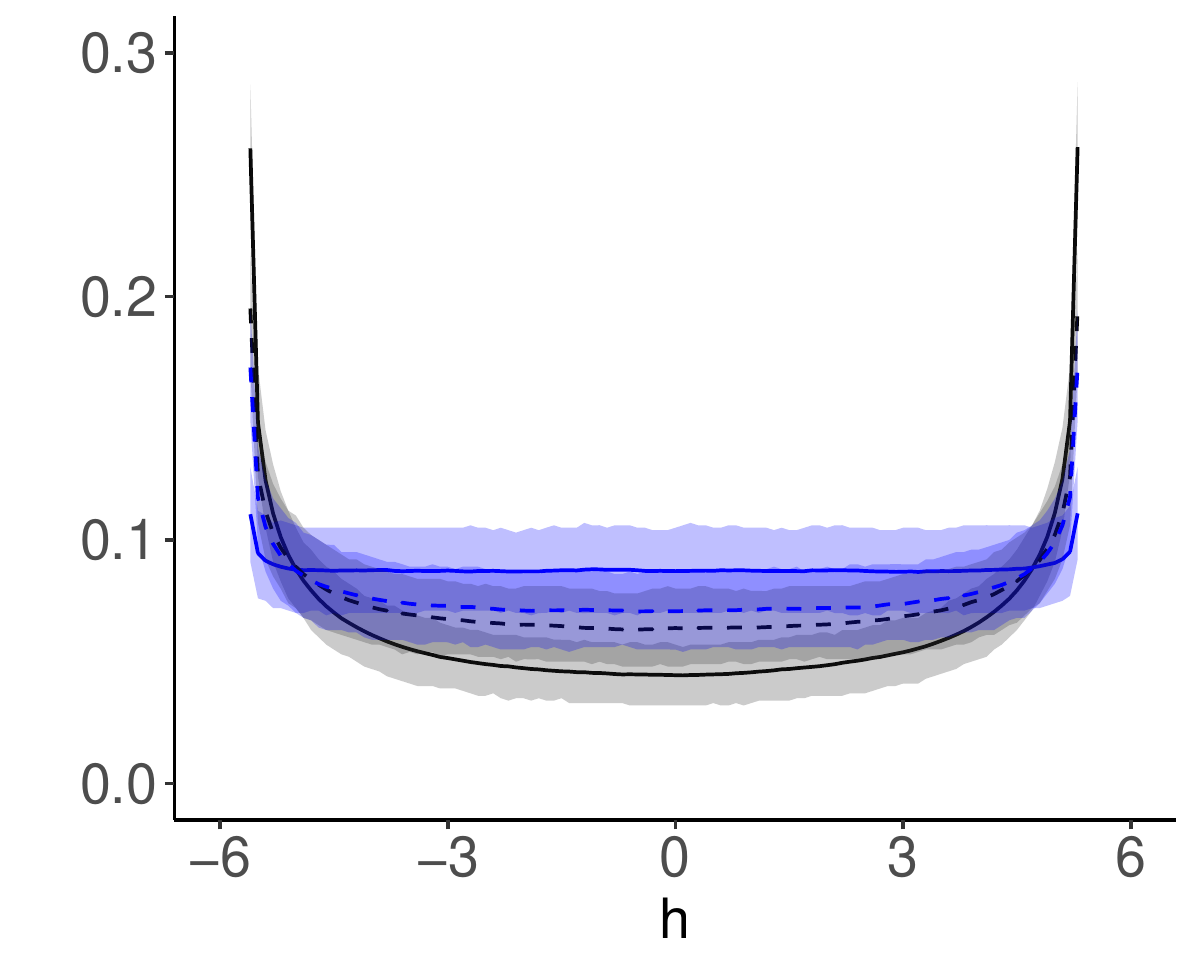} 
    \end{tabular}
    \caption{Left: Estimated probabilities of concurrent exceedances as a function of the distance $h$ for four stationary processes. Right: Estimated probabilities that the process $X$ exceeds $1$ on the interval $S_h = [h - 0.5,h+0.5]$ only as a function of its location $h \in [-5.5,5.5]$ in $S =[-6,6]$ when $X$ satisfies $\sup_{s \in \Ss} X(s) \geq 1 $ (right). Estimates are obtained by simulation of Gaussian processes (solid line) and generalize $\R$-Pareto process (dashed lines), for strong dependence (black) and weak dependence (blue).}
    \label{fig: boundary effect}
\end{figure}

\begin{theorem}\label{th: de novo}
Let $X$ be a stationary stochastic process with sample path on $C(\RR^d)$ satisfying Assumption \ref{assump:pi_decreasing} and $\Ss$ a compact set in $\RR^d$. For any convex subset $S_1 \subset \Ss$, the probability
$$
\Pr\{\sup_{s \in \Ss_1} X(s) \geq u , \sup_{s \in \Ss \setminus \Ss_1} X(s) \leq u \}, \quad u \in \RR,
$$
is a decreasing function of $\text{dist}(S_1,\dred{\partial}\Ss) = \inf_{s \in \Ss_1, s'\in \partial \Ss} \|s - s'\|$, where $\partial \Ss$ represents the boundary of $\Ss$.
\end{theorem}
Theorem \ref{th: de novo}, whose proof can be found in Appendix \ref{sec: proof de novo}, reveals that the supremum of stochastic processes is more likely to be located close to the boundary $\partial \Ss$ of $\Ss$.
Such boundary effect influences solutions to \eqref{eq: de nove princ simp} and, depending of the strength of dependence, favours points on, or close to $\partial \Ss$.
Figure \ref{fig: boundary effect} illustrates this phenomenon for two classical models of stochastic processes and different level of dependence; see Appendix \ref{app: simulation details} for simulation details.
We observe that exceedances above the threshold $u =1$ are more likely on and close to the boundaries. We also note that the boundary effect vanishes as soon as the set $\Ss_h$ reaches a distance greater than the dependence range: this phenomenon is observed for the second type a Gaussian process with weak dependence and, more generally, for any process for which (near)-independence is achieved for large distances.
Examining equation \eqref{eq: decomposition boundary} in light of Theorem \ref{th: de novo} shows a salient aspect of the optimization of sampling designs for monitoring extremes: it  yields non-trivial points patterns that results from a compromise between a boundary effect and maximization of inter-location distances.

One way to mitigate the boundary effect is to search for candidate locations not exclusively within $\Ss$ but in a region including $\Ss$ and its extension up to the effective range of dependence in all directions.
In this case, points inside $\Ss$ are, not surprisingly, systematically favoured as $\Pr\left\{X(s_1)\geq u \left|  \sup_{s \in \Ss} X(s) \geq u\right\}\right.$ in \eqref{eq: decomposition spacefilling} is maximal, and constant, for any $s_1 \in \Ss$.
In this case, the boundary effect is smoothed but the induced monitoring network is meant to be exposed to false positives. 
In the absence of boundaries, e.g., if $\Ss$ is a compact manifold, the sampling scheme simply cover the space as well as possible by maximizing the inter-locations distances corresponding to space-filling points sets, as a consequence of Assumption \ref{assump:pi_decreasing}.

We should stress once again that all the previous developments rely on Assumption \ref{assump:pi_decreasing} and simplification \eqref{eq: de nove princ simp}, and do not require hypotheses on the asymptotic tail behaviour of the process, as it is commonly done in extreme value theory.
Thus, any process whose dependence for finite threshold $u$ decreases with the distance exhibit such boundary effect.
These results however rely on strict stationarity, which is unrealistic for most applications.
When some a priori information is available, for instance from existing measurement stations, or from other indirect measurements, generalized $\R$-Pareto processes can be leveraged to find near-optimal sampling designs without assuming strict stationarity.

\subsection{Algorithm for optimal sequential sampling design}
\label{sec: alg near opt samp designs}
Solving \eqref{eq: samp princ} in its general form is intractable, so in practice we resort to a discretization of the region of interest.
The resolution of the grid is determined by the desired spatial precision of the solution and the accuracy of the probabilities estimates.
Even with such setting, tractability of the optimization cannot be ensured when searching for sets of size larger than $3$ because of the combinatorial nature of the problem.
We therefore suggest a sub-optimal, but tractable, method to solve \eqref{eq: samp princ}, i.e., we propose to search optimal sequential designs using Algorithm \ref{alg: sequential design}.
\begin{algorithm}
\SetAlgoLined
Input number of location $L$\;
Optional: initial set of points $S_{\rm samp}$ \;

\If{$S_{\rm samp} = \emptyset$}{
   Choose initial point preferably either at random or using geometrical information on $\Ss$\;
   }
\For{i from $L_{\rm samp} + 1$ to $L$}{
  Solve  $s_{\rm add} = \argmin{s \in S}\left|\Pr\left\{ \R(X) \geq u\right\} - \Pr\left\{ \R_{\{s_1,\dots,s_{i - 1}, s\}}(X) \geq u| \R(X) \geq u\right\}\right|$\;
  Set $S_{\rm samp} = S_{\rm samp} \cup \{s_{\rm add}\}$
 }
return $S_{\rm samp}$.
 \caption{Sequential algorithm for extremal sampling design.}
 \label{alg: sequential design}
\end{algorithm}
This procedure is general, applicable to any risk functional, and  for the special case of the supremum, provides an optimal set of locations as
\begin{align}
 &\Pr\left.\left\{ \max_{s \in \{s_1,\dots,s_L\}}X(s) \geq u\right| \sup_{s \in \Ss} X(s) \geq u\right\}\nonumber \\
  & = \Pr\left.\left\{ \max_{s \in \{s_1,\dots,s_{L-1}\}}X(s) \geq u\right| \sup_{s \in \Ss} X(s) \geq u\right\}\nonumber \\
 &  + \Pr(X(s_L)\geq u |  \sup_{s \in \Ss} X(s) \geq u) -\Pr\left( X(s_L) \geq u, \max_{s \in \{s_1,\dots,s_{L-1}\}}X(s) \geq u | \sup_{s \in \Ss} X(s) \geq u \right). \label{equ:space-filling and distance}
\end{align}
Indeed, the first term of the right hand side of \eqref{equ:space-filling and distance} is fixed given $s_1, \dots, s_{L-1}$, while the others depend only on $s_L$.
Thus, sequentially maximizing the left hand side of \eqref{equ:space-filling and distance} is equivalent to optimize the last two terms of the right side, and thus explicitly follows the principles derived from Theorem \ref{th: de novo}. 

\begin{figure}
    \centering
    \begin{tabular}{cccc}
        \includegraphics[width=0.2\textwidth]{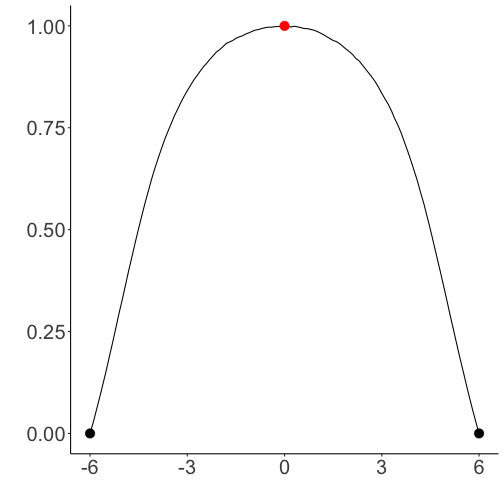} &  \includegraphics[width=0.2\textwidth]{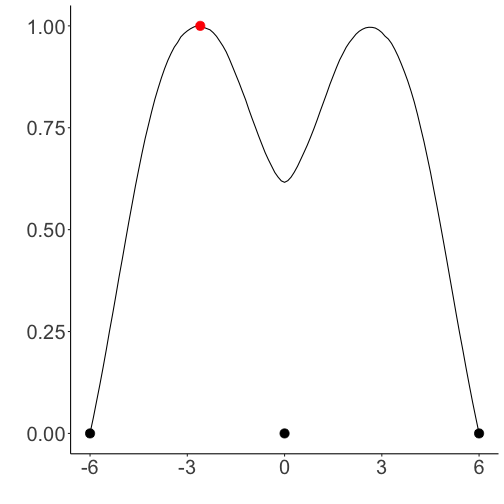}
         & \includegraphics[width=0.2\textwidth]{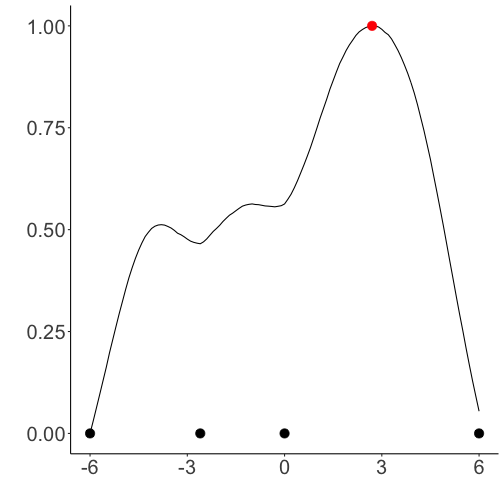} & \includegraphics[width=0.2\textwidth]{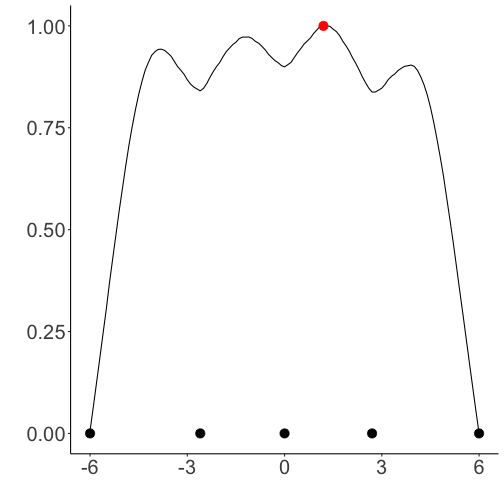} \\
          \includegraphics[width=0.2\textwidth]{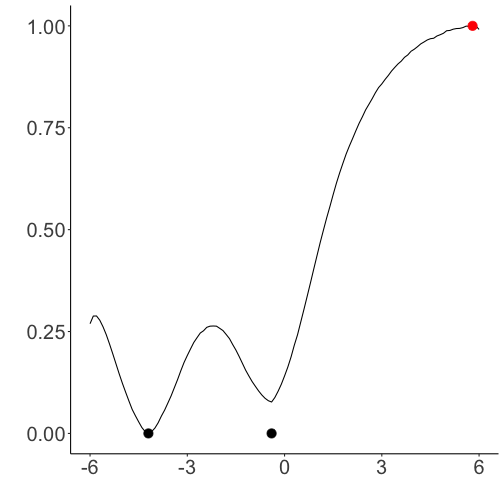} &  \includegraphics[width=0.2\textwidth]{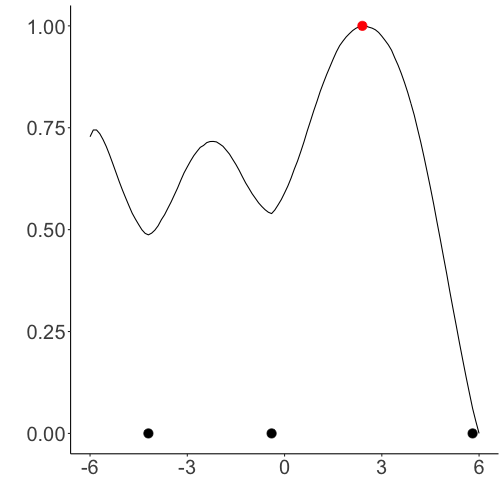}
         & \includegraphics[width=0.2\textwidth]{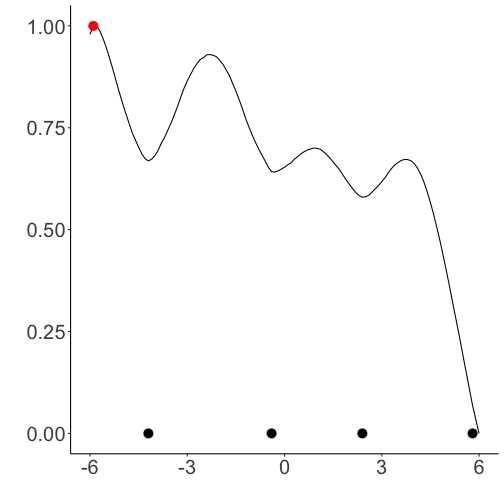} & \includegraphics[width=0.2\textwidth]{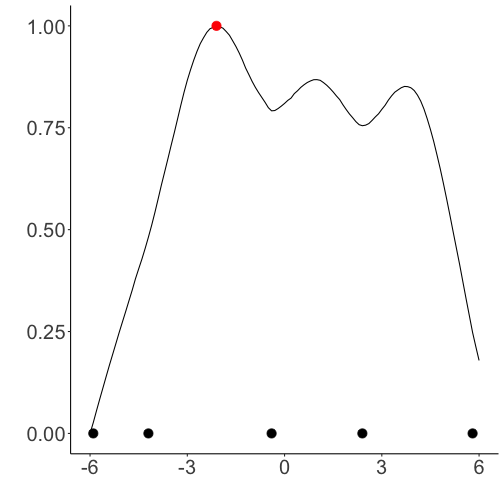}
    \end{tabular}
    \caption{Illustration of Algorithm \ref{alg: sequential design} for sequential addition of $4$ locations with boundary (top) and random (bottom) initialization. The curves represent the (conditional) probability of exceedance and are estimated empirically from $10^6$ simulation of an extremal stationary process satisfying $\sup_{s \in [-6,6]} x(s) \geq 1$ and scaled to varies between $0$ and $1$.}
    \label{fig: iterative sim}
\end{figure}

Figure \ref{fig: iterative sim} illustrates the procedure described in Algorithm \ref{alg: sequential design}; in this experiment, probabilities are estimated empirically using $10^6$ simulations of a weakly dependent $\R$-generalized Pareto process as in Figure \ref{fig: boundary effect} and rescaled to $0$ and $1$ for easier visualization. 
We consider two initialisation: first, two stations located on the boundary of $[-6,6]$ as prescribed by Theorem \ref{th: de novo} and, second, two stations selected at random.
In the latter, we observe that, as a consequence of the phenomenon described by Theorem \ref{th: de novo}, locations on the boundaries are quickly sampled.

In case where the interest lies beyond sequential design, Algorithm \ref{alg: sequential design} can be adapted to include a forward-backward steps such has in \citet{Zhang2011}. One could also consider $k-$look ahead strategies, that is the simultaneous inclusion of $k$ measurement stations. In the application described in Section \ref{sec: sihl application}, the implementation of a forward-backward step does not change significantly the selected locations.
\section{Extremal sampling designs using extreme value theory}\label{sec: evt model}
\subsection{Peaks-over-threshold analysis}\label{sec: gen Pareto theory}
Extreme value theory was first introduced for block maxima, describing the limit distribution of
$$
M_n = \max_{i = 1,\dots,n} X_i, \quad \text{as} \quad n \rightarrow \infty,
$$
where $X_i$ are independent and identically distributed random variables \citep[e.g.,][]{Fisher1928,gnedenko43}.
However, for more accurate risk quantification and optimal monitoring of extremes toward efficient early warnings of natural disasters, we need to study single events with potentially large values, i.e., tail distributions of stochastic processes. For this reason, we focus on an alternative view on extremes called peaks-over-threshold analysis \citep{Balkema1974,pickands1975,Davison1984}. 

Let $X$ be a random variable, then for any threshold $u<\inf\{x:F(x)=1\}$, under some mild conditions we can find sequences $a_n >0$ and $b_n$ such that
\begin{equation}\label{eq: gpd}
\left.
\begin{array}{ll}
n \Pr \left\{ \left(1 + \xi \frac{X - b_n}{a_n}\right)_+^{1/\xi} \geq x \right\}, & \xi \neq 0, \\
n \Pr \left\{  \exp\left(\frac{X - b_n}{a_n}\right)_+ \geq x \right\},  & \xi = 0,
\end{array}
\right\} \rightarrow x^{-1},  \quad n \rightarrow \infty,
\end{equation}
where $(\cdot)_+ = \max(\cdot,0)$ .
In practice, \eqref{eq: gpd} means that the conditional distribution of exceedances over a high threshold can be approximated by a {generalized Pareto} distribution.
The parameter $\xi$, called the {tail index}, determines the strength of the tail and its support: for $\xi > 0$, $x \geq u$ and the tail decay is polynomial (Fr\'echet),
for $\xi = 0$, $x \geq u$ and the tail has an exponential decay (Gumbel), and finally for $\xi < 0$, $x \in (u;u - \sigma/\xi)$ and we retrieve a polynomial tail decay (Weibull).
This result has been generalized to a multivariate setting \citep{{Rootzen2006}, Rootzen2018, {Rootzen2017}}, and more recently to functions \citep{DeFondeville2017}.

Again, let $\Ss$ be a compact subset of $\RR^d$ representing the region under consideration and $C(\Ss)$ denotes the space of real-valued continuous functions over $\Ss$. We now consider that $X$ refers to a stochastic process with sample path in $C(\Ss)$.
In a univariate context, it is straightforward to define the notion of exceedance, but for functions, it needs to be carefully introduced.
Following \citet{Dombry2013} and \citet{Fondeville2017}, we consider a functional $\R: C(\Ss) \rightarrow \RR$, called a risk functional, which computes a univariate summary of the stochastic process $X$ and defined an $\R$-exceedance as an event $\{\R(X) \geqslant u\}$ where $u \geqslant 0$ is a threshold of choice.

We now suppose that there exist sequences of functions $a_n \in C(\Ss,(0,\infty))$ and $b_n \in C(\Ss)$ and a scalar $\xi \in \RR$ such that
\begin{equation}\label{eq: reg var}
n \Pr \left\{ \left(1 + \xi \frac{X - b_n}{a_n}\right)^{1/\xi} \in \cdot \right\} \rightarrow \Lambda(\cdot),
\end{equation}
where  $\Lambda$ is a non-degenerate  measure on $\Set = C(\Ss,(0,\infty)) \setminus \{0\}$.
Equation \eqref{eq: reg var} is a natural extension of the univariate condition \eqref{eq: gpd} for convergence of the tail distribution towards a generalized Pareto distribution and implies that
\begin{equation}\label{eq: gen r Pareto}
\Pr\left.\left[ \frac{X - b_n}{a_n} \in (\cdot) \right| \R\left\{\frac{X - b_n}{a_n}\right\} \geqslant 0\right] \rightarrow \Pr\{P \in (\cdot)\}, \quad n \rightarrow \infty,
\end{equation}
where $P$ is a generalized $\R$-Pareto process; detailed conditions on the functional $\R$ can be found in \citet{DeFondeville2017}.

The generalized $\R$-Pareto process $P$ is parametrized by a tail index $\xi$ function and a limit measure $\Lambda$, respectively characterizing the regime of marginal tail decay and extremal dependence.
This class of processes is a natural extension of univariate peaks-over-threshold analysis to functions. In particular, for any $s \in \Ss$ and sufficiently high threshold $u \geqslant 0$,
$$
\Pr\{P(s) \geqslant u + x \mid P(s) \geqslant u\} = 1- F_{\xi,\sigma(u)}(x),
$$
where $\sigma(u) = 1 + \xi u$ and $1-F_{\xi,\sigma(u)}$ is the survival function of a generalized Pareto random variable.
However, we request that the measure $\Lambda$ is non-zero on the space of continuous functions on $\Ss$, which restricts the model to one particular case of extremal dependence, namely asymptotic dependence.
This implies that the strength of the dependence decreases but stabilizes as the intensity of the event increases, i.e., there exists for each location in $s \in \Ss$ a neighbourhood for which points are positively associated with $X(s)$ independently of the marginal intensity.

Generalized $\R$-Pareto processes are constructed using the representation
\begin{equation}\label{eq: pole repr}
P = \frac{\{RW/\R(W)\}^\xi - 1}{\xi},
\end{equation}
where $R$ is a univariate unit Pareto random variable and $W$ is a stochastic process on the unit $L_1-$sphere $\{x \in C(\Ss): \|x\|_1 = 1\}$ whose distribution is determined by $\Lambda$ and models the dependence of $P$.
Equation \eqref{eq: pole repr} is a key component to sampling algorithms as the ones described in \citet{DeFondeville2017}.
In general, the process $W$ can take any kind of distribution as soon as $\Esp\{W(s)\} = 1$ for all $s \in \Ss$.

Assuming that the tail index $\xi$ is homogeneous over $\Ss$ is common in environmental applications \citep[e.g.,][]{Ferreira2012a, Engelke}, especially when the region of interest is small relative to the scale of the process under study.
In this case, the tail index $\xi$ characterizes not local tail behaviors but the tail regime of the physical process itself, e.g., a specfic type of rainfall.
Visual diagnostics, such as qq-plots, allows to assess the reliability of such assumption.
In general, the latter can be relaxed, but then, using an asymptotically justified model requires to define exceedances on the re-normalized process $Y = \{1 + \xi(X - b_n)/a_n\}^{1/\xi}$, with the risk to lose any potential physical interpretation.

To model extremal dependence, a very convenient model for $W$ is a stationary log-Gaussian random field, for which extremal dependence is simply summarized by
\begin{equation}\label{eq: extremogram}
\rho(h) = \Pr\{ P(s+h) \geqslant u \mid P(s) \geqslant u\} = 2 \left(1 - \Phi\left[\left\{\frac{\gamma(h)}{2}\right\}^{1/2}\right] \right),
\end{equation}
where $u$ is a sufficiently large threshold, $\gamma$ is the semi-variogram of the underlying Gaussian process and $\Phi$ is the distribution function of a standard normal random variable.
The function $\rho$ is called the extremogram \citep{Davis2009}, or the $\chi$ coefficient in a multivariate setting \citep{Trust2017}, and measures pairwise extremal dependence.
Indeed, for heavy-tailed processes, classical measures of dependence such as a covariance function does not necessarily exists, and thus we require alternatives such as the extremogram.
This model is convenient as it allows to leverage existing Gaussian dependence models from the literature on spatio-temporal statistics to drive the extremal dependence of $P$.

In practice, following asymptotic convergence \eqref{eq: gen r Pareto}, the distribution of functional exceedances can be approximated by a generalized $\R$-Pareto process, i.e., 
\begin{equation}\label{eq: pareto approx}
\Pr\left.\left[ \frac{X - b_n}{a_n} \in \cdot \right| \R\left\{\frac{X - b_n}{a_n}\right\} \geqslant 0\right] \approx \Pr\{P \in \cdot\}
\end{equation}
where the functions $a_n$ and $b_n$ are also unknown and need to be estimated.
When the risk functional is linear, i.e., the condition in \eqref{eq: pareto approx} simplifies to exceedances of $\R(X)$ above $u = b_n$, a large $n$ is equivalent to a large threshold $u \in \RR$.
In this case, the random variable $\R(P)$ follows a generalized Pareto distribution with tail index $\xi \in \RR$, scale $a_n >0$ and location $b_n \in \RR$.

Generalized $\R$-Pareto processes offers a simple and flexible solution to model extremes of random fields. It is the main building block of the methodology proposed here to create, or enlarge, network of monitoring stations, for optimal monitoring of extreme events.

\subsection{A flexible model for station networks}
\label{sec: design with a priori}
We now propose a model to find an optimal sequential sampling design in the sense of \eqref{eq: samp princ} when some a priori information of the physical process under consideration is available.
For the Sihl river, as presented in Section \ref{sec: intro}, three monitoring stations are installed inside the basin and radar observations cover the whole region, providing information about the intensity and the structure of extreme rainfall in the region.
The goal of such an analysis is to leverage all available data and propose guidelines as to where new stations should be installed for optimal monitoring of extremes.

More precisely, we suppose that, for a risk function $\R$, the distribution of $\R$-exceedances of the process $X$ over the threshold $u \in \RR$ can be approximated by a generalized $\R$-Pareto process $P$ as in \eqref{eq: pareto approx}.
In this case, the process is allowed to have both non-stationary marginal tail distributions and potentially non-stationary extremal dependence.
We also suppose that it is possible to estimate the parameters of $P$ from the available data, i.e, that we have estimates of the functions $\hat{a}_n$, $\hat{b}_n$, $\hat{\xi}$ and of the dependence structure of $W$.
In practice, this implies imposing a marginal parametric model such as
$$
\left.\begin{array}{ll}
a_n(s) & = f\{y(s)\} > 0, \\
b_{\dred{n}}(s) & = g\{y(s)\} \in \RR, \\
\xi(s) & = \xi \in \RR,
\end{array}\right\} s \in \Ss,
$$
where $y$ refers to a vector of covariates, such as locations $s \in S$ or any field available throughout $\Ss$ like altitude or any spatially dense data set such as radar acquisitions.
As mentioned in Section \ref{sec: gen Pareto theory}, assuming constant $\xi$ is common in environmental applications and is necessary to ensure that the risk can be defined on the original scale.
We also suppose that the dependence of $W$ is parametrized by a vector $\theta_W$ of parameters.
For instance, if $W$ is log-Gaussian, $\theta_W$ refers to the parameters of the semi-variogram function $\gamma$, which can be allowed to be non-stationary over space.
Estimation procedures for $a_n$, $b_n$, $\xi$ and $\theta_W$ can be found in \citet{DeFondeville2017} .

We also assume that the process $X$ is observed at a set of locations $S_{\text{obs}} = \{s_1,\dots, s_{L_{\text{obs}}}\} \subset \Ss$; the latter could be empty if no station measures are available and alternative sources of data can be leveraged.
We denote by $L_{\rm samp}$ the number of new sites, i.e., the size of the network extension. 
This parameter is usually determined by practical constraints and available budget.
With a non-stationary marginal behaviour, and potentially dependence, finding an optimal sampling design is not trivial, so, as in \ref{sec: alg near opt samp designs}, we approximate numerically equation \eqref{eq: samp princ} with a finite number of potential candidate sites, i.e., we use a grid over $\Ss$ of size $L_{\rm grid}$, whose size is determined by the resolution of the covariates $y$.
With this setting, the number of sets of candidate sites grows exponentially with $L_{\rm samp}$, getting quickly intractable for $ L_{\rm samp} > 2$.
As detailed in Section \ref{sec: alg near opt samp designs}, we propose a sequential procedure to solve \eqref{eq: samp princ}; this provides a reasonable approximation of the solution while being computationally efficient. A sequential solution also reflects fields practices as in general new weather stations are installed sequentially, as this is a time consuming and costly process.
The procedure is summarized by Algorithm \ref{alg: optimal sampling} and combines an empirical estimator of the $\R$-exceedance probabilities with simulations from the fitted generalized $\R$-Pareto process.
For the supremum functional, Algorithm \ref{alg: optimal sampling} provides an exact solution to sequential designs as it was proved in Section \ref{sec: alg near opt samp designs}. Note that, for instance, if the risk functional is linear and $u = \R(b_n)$, then we can simply set $I = 1$, as by definition, all the elements of the simulation set $P_n$ satisfies $\R(P_n) \geq u$.

\begin{algorithm}
\SetKwInput{KwInput}{Input} 
\SetAlgoLined
\KwInput{$S_{\text{obs}}$, $S_{\rm grid}$, $u$, $a_n$, $b_n$, $\xi$ and $\theta_W$\;
number of desired new sampling points $L_{\text{samp}}$, number of simulations $N$\;}

Simulate $N$ generalized $\R$-Pareto processes $P_n$ with parameters $(\xi,\theta_W)$\;
Compute $I = N^{-1} \sum_{j = 1}^N 1\{\R_{S_{\rm grid}}(a_nP_j + b_n) \geq u\}$\;
Set $S_{\rm samp} = \emptyset$\;
  \For{l from $1$ to $L_{\rm samp}$}{
   \For{k in $1$ to $L_{\rm grid}$}{
     Set $R[k] =|I - N^{-1} \sum_{j = 1}^N 1\{\R_{s_k \cup S_{\rm samp} \cup S_{\rm obs}}(a_nP_j + b_n) \geq u\}|$\;
   }
   Set $S_{\rm samp} = S_{\rm samp} \cup s_{k_{\max}}$ where $k_{\max} = \argmax{k} R[k]$
 }
 
return $S_{\rm samp}$.
 \caption{Algorithm for near-optimal sequential sampling design to monitor extremes}
 \label{alg: optimal sampling}
\end{algorithm}

\section{Optimal network for the Sihl river basin}\label{sec: sihl application}
To study rainfall in the region of Zurich, we can rely on the existing network of weather stations.
The Sihl river basin itself contains only three monitoring stations. For improved model inference, we estimate the model using the measurements available in the coloured region in Figure \ref{fig: basin and stations}, which includes up to $15$ monitoring stations.
For all these sites, MeteoSwiss provides hourly mean rainfall with measurement from January $1^{\rm st}$ $2013$ to March $2020$ for half of them, while the $7$ others have records starting between $2014$ and $2016$.
Thus the times series have between $30 000$ and $62 000$ measures.
One station outside the river basin has been installed in $2019$ and therefore does not include enough measurements to be safely included.

To estimate a model for a generalized $\R$-Pareto process, we can also leverage radar products. More precisely the CombiPrecip data set \citep{Sideris2014a,Gabella2017,Panziera2018}  produced by MeteoSwiss provides estimates of hourly rainfall accumulation since $2013$ on dense grid of $1$km resolution.
Earlier measurements are also available but inconsistent with recent acquisition due to hardware and processing changes in 2013.
The Sihl river basin is orographically homogeneous and located at a reasonable distance from the radar, so the estimated rain fields can be assumed to be fairly homogeneous and free from processing biases.
These measures, as the result of some processing, are however not reliable by themselves, i.e., their link with station measurement is unknown.
We however suppose that the radar estimates present spatial variations similar to the true underlying rain field.
Thus, we use radar measurements, first, as covariate to extrapolate the tail marginal model away from existing weather stations and, second, to estimate the model for extremal dependence. 

\subsection{Marginal model}\label{sec: marginals}
\begin{figure}
\begin{center}
\includegraphics[width=0.4\textwidth]{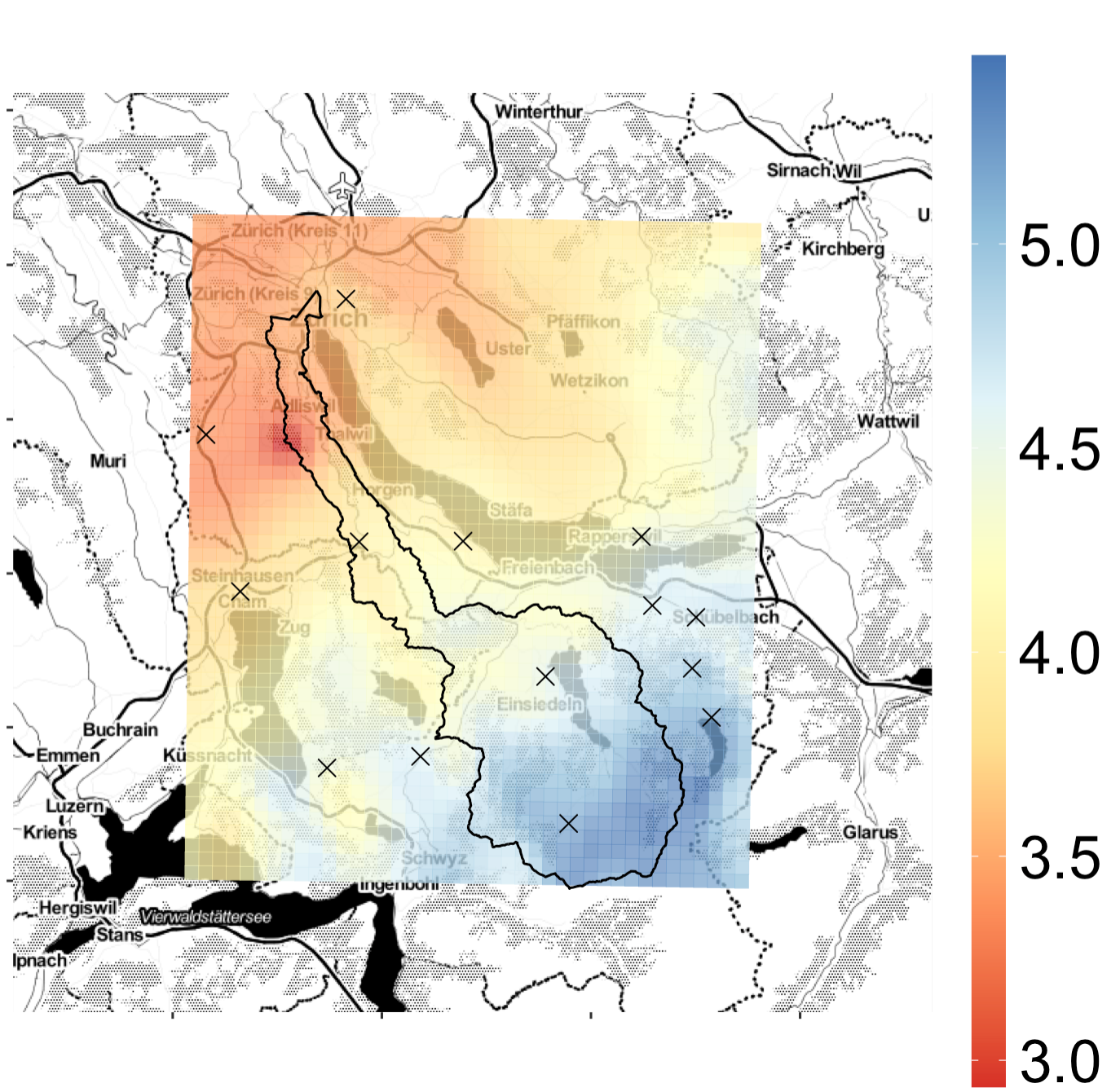}
\end{center}
\caption{Estimated location model $\hat{b}_n$. The estimate is obtained using a least squares algorithm with radar based hourly mean rainfall as covariate. The solid black line delimits the Sihl river basin and black crosses represent the location of weather stations.}\label{fig: modelled location}
\end{figure}

In order to use Algorithm \ref{alg: optimal sampling}, we need first to estimate the functions $a_n$, $b_n$ and the tail index $\xi \in \RR$. To do so, we assume the following parametric model
\begin{equation}\label{eq: parametric model}
\left.\begin{array}{ll}
a_n(s) & = a \in (0,\infty), \\
b_n(s) & = b_1 + b_2 \times y(s), \\
\xi(s) & = \xi \in \RR,
\end{array}\right\} s \in \Ss,
\end{equation}
where $y$ is a covariate derived from radar measurements.
This model has been chosen following a preliminary analysis, where generalized Pareto distributions were fitted independently for each of the $14$ stations and various thresholds: we found both the scale and tail index parameters to be homogeneous across $\Ss$ and $\xi$ to be stable around quantiles $0.995$ of the observational series; this represents between $150$ and $300$ exceedances per station, and about $2.5\%$ to $5\%$ of wet days.
We explored several candidates for the covariate $y$, including local quantiles at different levels, and average rainfall accumulation, both for wet days only and overall series; we found the later to have about $95\%$ correlation with empirical $0.995$ quantiles from the station measurements. 
We thus chose the later as covariate in the parametric model \eqref{eq: parametric model}; its spatial variation is displayed in Figure \ref{fig: basin and stations}. 
Then using a least squares algorithm, we obtain $\hat{b}_1 = 1.14(0.34)$ and $\hat{b}_2 = 20.8 (2.1)$; numbers in the brackets refer to estimated standard deviations.  
The corresponding model $\hat{b}_n$ is displayed in Figure \ref{fig: modelled location} and gives values higher than the corresponding empirical quantiles of the radar product.
We then use independent likelihood to estimate the common scale $\hat{a} = 1.87 (0.05)$ and tail index $\hat{\xi} = 0.33 (0.02)$.
The fit is assessed by visual inspection of the QQ-plots given in Appendix \ref{app: qqplots}, which is particularly convincing as it accommodate about $3200$ exceedances from $14$ stations with only $4$ parameters.
It is not surprising that such a simple model presents a good level of performance as the region under consideration is relatively small, approximately $2000$km$^2$, with limited variations in altitude and is thus fairly homogeneous from an hydrological perspective.

\subsection{Dependence model}\label{sec: dependence sihl}
For generalized $\R$-Pareto processes, extremal dependence is fully characterized by the angular component $W$, for which we propose to use a log-Gaussian process. More precisely,
\begin{equation}\label{eq: log-Gauss pareto}
W(s) = \frac{\exp\left\{ G(s) - \sigma^2(s) /2\right\}}{\| \exp\left( G - \sigma^2 /2\right)\|_1},
\end{equation}
where $\sigma^2(s)$ is the variance at location $s\in \Ss$ of a zero mean Gaussian process $G$.
We further assume that $G$ has stationary increments, i.e., its semi-variogram function $\gamma(s,s') = \Var\{G(s') - G(s)\}$ is a function of the distance $h = s' - s$.
We remind that in this case, the pairwise extremal dependence, as summarized by the extremogram, is linked through the closed form \eqref{eq: extremogram} to $\gamma$, for which we impose parametric model. 
We choose \citep{Schlather2017}
\begin{equation}\label{eq: vario model}
\gamma(h) = \frac{ (1 + \|Ah/\lambda\|^\alpha)^{\beta / \alpha - 1} }{2^{\beta / \alpha} - 1}
\end{equation}
where $0 <\alpha < 2$ and $\beta < 2$ respectively drive the smoothness and long range behaviour of the process, $\lambda >0$ is a scale parameter and $A$ is a geometrical anisotropy matrix
$$
A = \left[\begin{array}{cc}
   \cos \delta  & -\sin \delta \\
   \kappa \sin \delta  &  \kappa \cos \delta
\end{array}\right],
$$
with $\kappa > 0$ and $\delta \in (0, \pi / 4)$.
The model \eqref{eq: vario model} is particularly attractive for modelling extremal dependence as, when $\beta < 0$, the semi-variogram function is bounded, while when $0< \beta < 2$, $\gamma(h) \rightarrow \infty$ as $h \rightarrow \infty$, meaning that the process tends to be independent for increasingly large distances.

The extremogram can be estimated from the observations with
\begin{equation}\label{eq: extremogram estimator}
\hat{\rho}(h) = \frac{\sum_{i = 1}^n 1\{X_i(s + h) \geq u_q(s + h), X_i(s) \geq u_q(s)\}}{1\{\sum_{i = 1}^n X_i(s) \geq u_q(s)\}}
\end{equation}
where $u_q(s)$ is the $q^{th}$ quantile of $X$ at location $s$, $q \in (0,1)$ close enough to $1$.
We apply equation \eqref{eq: extremogram estimator} on radar rainfall measurements with quantile level $q = 0.995$.
We then fit a parametric model for the semi-variogram function $\gamma$ using a least squares procedure.

\begin{figure}
\begin{center}
\begin{tabular}{cc}
\includegraphics[width=0.4\textwidth]{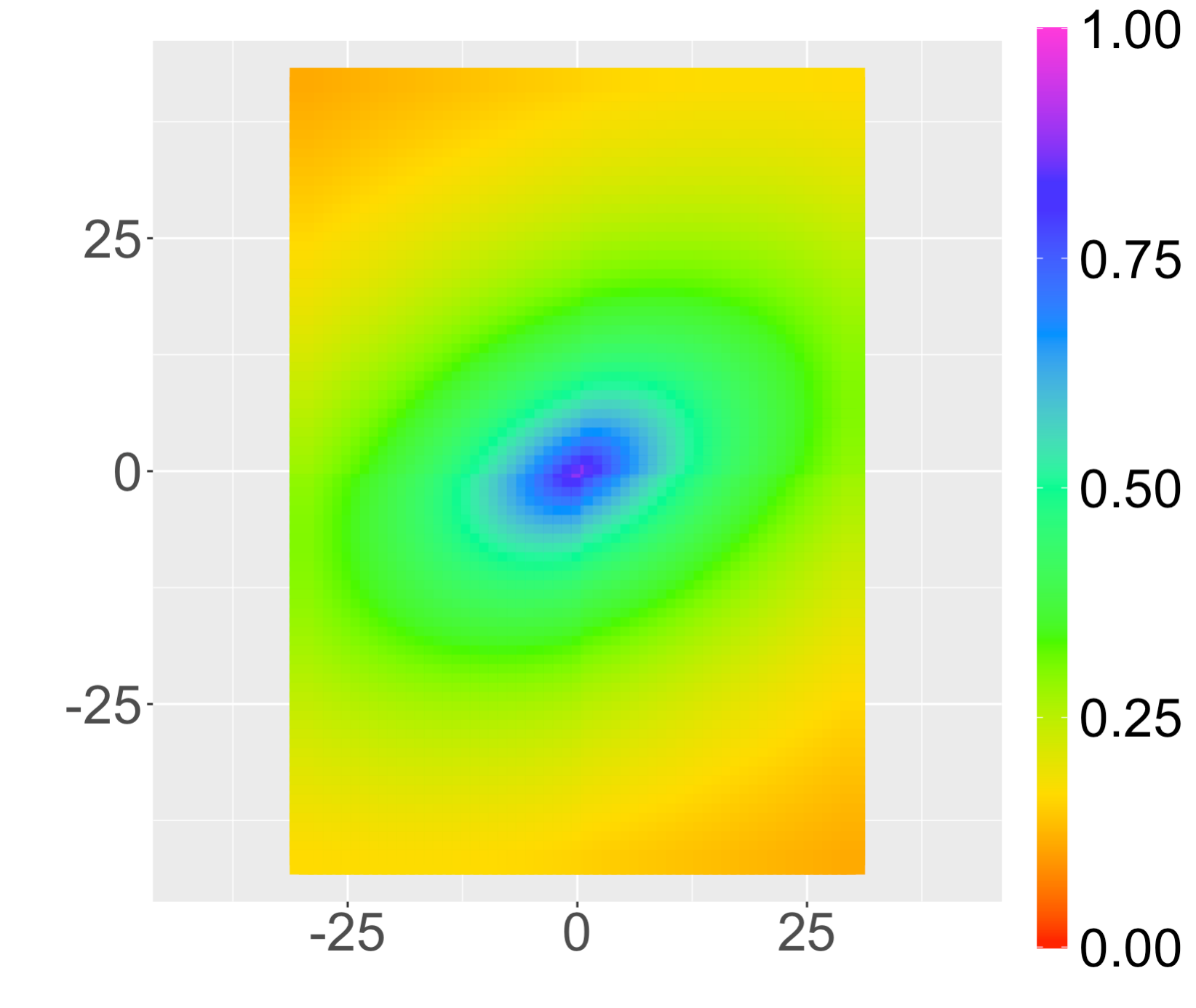} & \includegraphics[width=0.4\textwidth]{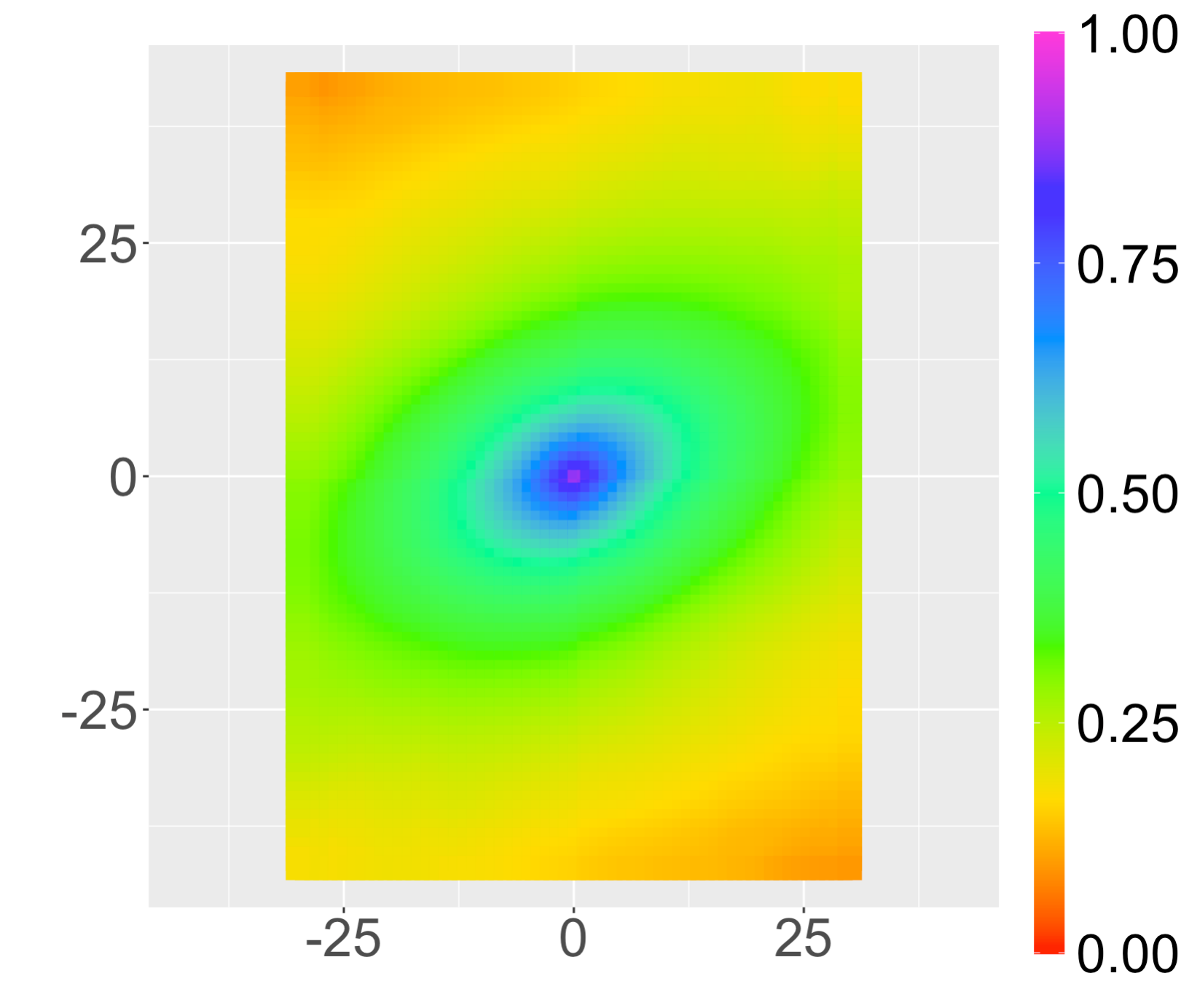}\\
Empirical Estimates & Fitted Model
\end{tabular}
\end{center}
\caption{Empirical (left) and modelled (right) pairwise dependence estimated from radar rainfall measurements.}\label{fig: est dependence model}
\end{figure}

Figure \ref{fig: est dependence model} displays the empirical estimates against the estimated model.
We observe a rather strong dependence up to $20$km especially in the north-west direction, which drops quite quickly after $25$km.
These estimates are consistent with the observation that at such a fine scale, rainfall events tends to be localized in regions with diameters of no more than few dozens of kilometers.

\subsection{New measurement stations in the Sihl Basin}
\begin{figure}
\begin{center}
\begin{tabular}{cc}
\includegraphics[width=0.4\textwidth]{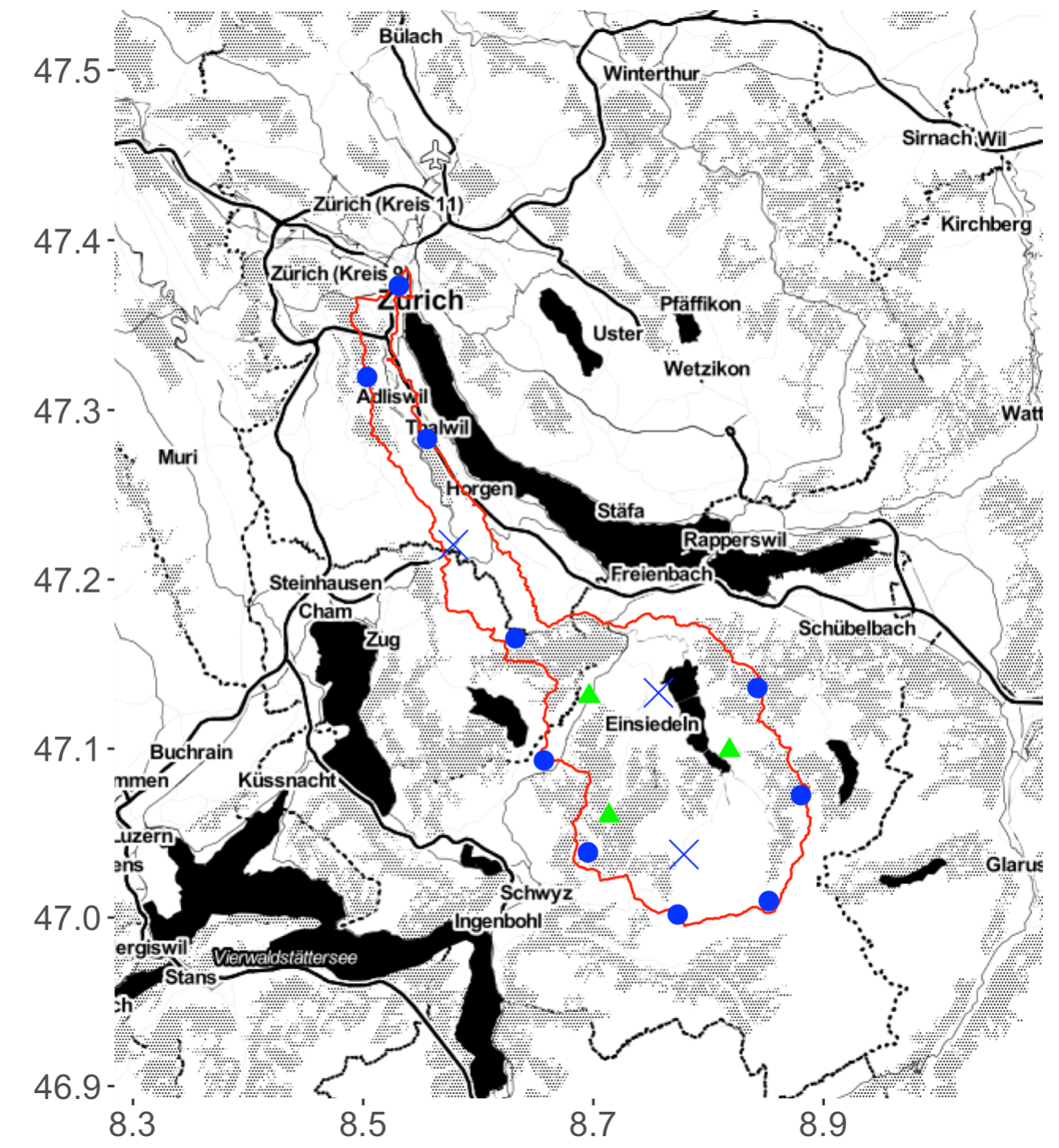} &
\includegraphics[width=0.4\textwidth]{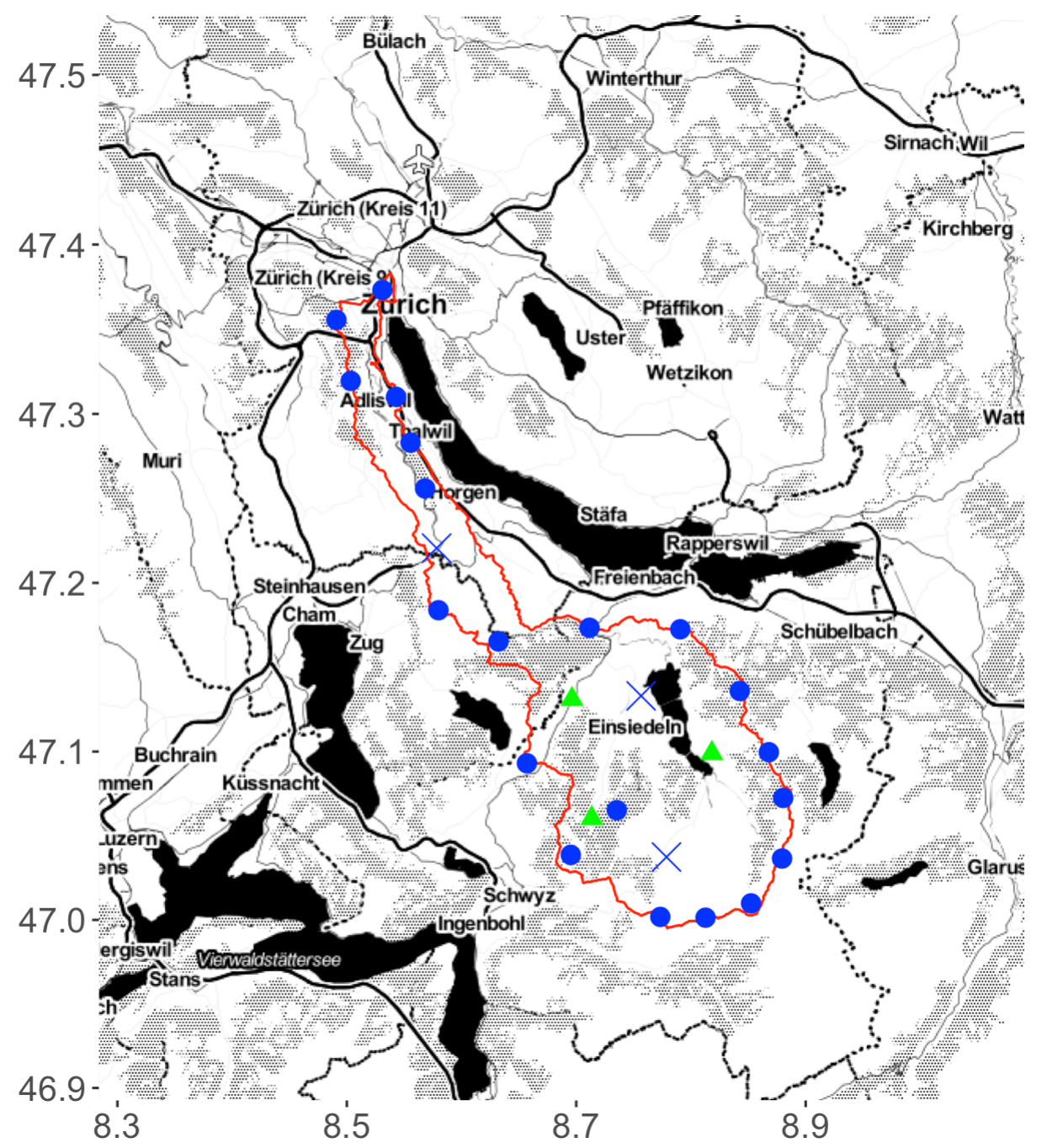}
\end{tabular}
\end{center}
\caption{Locations of $10$ (left) and $20$ (right) new weather stations for optimal monitoring of extreme rainfall inside the Sihl river basin with sequential sampling designs. Blue crosses: Existing stations; blue dots: new stations; plain red line: basin limit; green triangles: existing manual measurements stations.}\label{fig: new stations 10}
\end{figure}

Algorithm \ref{alg: optimal sampling} relies on the generation of large samples of generalized $\R$-Pareto processes, i.e., simulations of the angular component $W$, for which a detailed algorithm can be found in Appendix \ref{app: algo sim log-Gaussian}.
The model's output for the risk functional $\R(X) = \sup_{s\in \Ss}X(s)$ with threshold $u = 20$mm and $10^5$ simulations are displayed in Figure \ref{fig: new stations 10} for sequential extensions by $10$ and $20$ stations.

We also observe that as the number of sites increases, recommended locations tend to be located on the boundaries of the region, consistently with the results of Section \ref{sec: properties extr sampling}.
The very first station to be added is located in Zurich at the train station, however its relevance is limited as a station is already installed few kilometers away from the recommended location, as it can be seen in the Figure \ref{fig: basin and stations}.
The algorithm could be adapted to account for this phenomenon by extending the region by a few kilometers in all directions, mitigating the boundary effect but allowing for false positive when the network is used for early warning. 
The Sihl river basin also includes three stations that are manually operated by MeteoSwiss but can unfortunately not be leveraged for statistical analysis due to the temporal irregularity of the measurements. 
In Figure \ref{fig: new stations 10}, we can see that Algorithm \ref{alg: optimal sampling} suggests to automate one of them for the $20$ stations scenario.

\section{Discussion}\label{sec: discussion}

In this work, we propose a new principle for sampling designs aiming to achieve optimal monitoring of extremes.
We study the theoretical properties of the induced point patterns when monitoring the supremum of stationary stochastic processes whose dependence is assumed to decrease with distance.
We obtain sampling designs resulting from a compromise between inter-location distance maximization and a boundary effect whose influence is determined by the strength of dependence. 
We also propose a tractable algorithm, relying on sequential designs, to approximate our criterion for optimal sampling.
When data is available in the region of interest, as it is the case of Sihl river basin, we propose a model, based on generalized $\R$-Pareto processes, allowing us to propose recommendations for possibly extending the existing network by $10$ and $20$ stations.

Compared to other approaches, such as Gaussian processes, the proposed methodology allows a principled modelization of the tail of distribution, by using an asymptotically justified framework. The dependence structure that can be modelled is quite flexible, and could be made non-stationary using techniques such as \citet{Fuglstad2015} or \citet{Fouedjio2015}, when a sufficiently large number of stations is already in operation.
Our procedure can be efficiently implemented using the algorithm described in the Section \ref{sec: alg near opt samp designs}.
This contribution should also prove to be useful in a wide variety of settings.
For instance, \citet{Chang2007}, due to the absence of better and more specialized alternatives, resorted to use a multivariate Gaussian-Inverse Wishart hierarchical Bayesian distribution to monitor air pollution.
However, as they say, ``the primary role -- of air pollution monitoring networks -- is the detection of noncompliance with air quality standards based on extremes designed to protect human health'', and thus our model, which is tailored for extremes monitoring, would be particularly relevant in this context.

Finally, the proposed model for the Sihl river relies on generalized $\R$-Pareto processes, for which the strength of extremal dependence is `frozen' above a given level of intensity: the model is thus conservative by nature and extremal dependence might be over-estimated above the quantile of reference.
Asymptotic models cannot accommodate for decreasing trends of dependence with intensity that is observed in multiple environmental applications and sub-asymptotic alternatives for functional peaks-over-threshold should then be considered. However, only very few of them exists \citep{Davison2011,Huser2017,Huser2019} and would either not be tractable or sufficiently realistic in the application considered in this work.

\bibliography{library}
\bibliographystyle{apalike}

\appendix
\section{Simulation parameters for Figure \ref{fig: boundary effect}}\label{app: simulation details}
In Figure \ref{fig: boundary effect}, estimates are obtained by estimating of empirical probabilities using $1000$ samples of each processes.
The procedure in repeated $1000$ times to also obtain $95\%$ confidences intervals.
Both Gaussian and generalized $\R$-Pareto processes are defined on $S = [-6;6]$ and simulated over a regular grid with lag $0.1$.
For the Gaussian processes, we use semi-variogram function
$$
\gamma_G(h) = \sigma [1 - \exp\{ ( h/\lambda)^\alpha\}], \quad h > 0,
$$
with $\sigma = 2$, $\alpha = 1.5$, $\lambda = 100$ for the strong dependence case and $\lambda = 1$ for the weakly dependent case.
Generalized $\R$-Pareto processes are constructed using model \eqref{eq: log-Gauss pareto}, for which we use:
\begin{itemize}
    \item for the strong dependence case,
$$
\gamma_W(h) = \sigma [1 - \exp\{ ( h/\lambda)^\alpha\}], \quad h >0,
$$
with $\sigma = 2$, $\alpha = 1.5$, and $\lambda = 10$;
    \item for the weak dependence case,
$$
\gamma_W(h) =  ( h/\lambda)^\alpha, \quad h > 0
$$
with  $\alpha = 1.5$, and $\lambda = 2.5$.
\end{itemize}

\section{Proof of Theorem \ref{th: de novo}}\label{sec: proof de novo}
Suppose that $X$ is a stationary process on $C(\RR^d)$ with $d > 0$.
For simplicity, we prove Theorem \ref{th: de novo} for $d = 1$ but generalization to $d > 1$ is can be done using similar arguments.
We study the compact subset $[0,\Ss]$ with $ S > 0$ and let $\dred{\tilde{s}} \in [0,\Ss]$ and consider a threshold $u \in \RR$ such that $\Pr\{X(s) \geq u \} > 0$. 
Let $h \in H =  \dred{]0},\Ss-\tilde{s}]$ and define the function $\mu: H \rightarrow\RR$ such that
\begin{align*}
\mu(h) &  = \Pr\{\sup_{s \in [h,\dred{\tilde{s}}+h]} X(s) \geq u, \sup_{s \in [0,\Ss] \setminus [h,\dred{\tilde{s}}+h]} X(s) \dred{<} u\} \\
& = \Pr\{\sup_{s \in [h,\dred{\tilde{s}}+h]} X(s) \geq u\} - \Pr\{\sup_{s \in [h,\dred{\tilde{s}}+h]} X(s) \geq u, \sup_{s \in [0,\Ss] \setminus [h,\dred{\tilde{s}}+h]} X(s) \geq u\}
\end{align*}
and we aim to prove that $\mu(h)$ is decreasing if $h < (\Ss -\dred{\tilde{s}})/2$ and increasing when $h > (\Ss -\dred{\tilde{s}})/2$, i.e., $\mu(h)$ is a decreasing function of the distance to the boundaries of $d([h,h+\dred{\tilde{s}}],\partial S) = \inf_{s\in [h,h+\dred{\tilde{s}}]} \inf_{s'\in \partial S} d(s,s')$.

For any $h' \in [0,\frac{S-(\dred{\tilde{s}} + h)}{2}]$, using the stationarity of $X$, we have
\begin{align*}
    \mu(h + h') &  = \Pr\{\sup_{s \in [h + h',\dred{\tilde{s}}+h +h']} X(s) \geq u, \sup_{s \in [0,\Ss] \setminus [h + h',\dred{\tilde{s}}+h+h']} X(s) \leq u\},\\
    & = \Pr\{\sup_{s \in [h,\dred{\tilde{s}}+h]} X(s) \geq u, \sup_{s \in [-h',\Ss - h'] \setminus [h,\dred{\tilde{s}}+h]} X(s) \leq u\},\\ 
    & = \Pr\{\sup_{s \in [h,\dred{\tilde{s}}+h]} X(s) \geq u, \sup_{s \in [-h',0] \setminus [h,\dred{\tilde{s}}+h]} X(s) \leq u, \sup_{s \in [0,\Ss -h'] \setminus [h,\dred{\tilde{s}}+h]} X(s) \leq u\},\\
    & = \Pr\{A(h,h'), \sup_{s \in [-h',0] \setminus [h,\dred{\tilde{s}}+h]} X(s) \leq u\}, \\
    & = \Pr\{A(h,h')\} -  \Pr\{A(h,h'), \sup_{s \in [-h',0]} X(s) \geq u\}
\end{align*}
where
$A(h,h') = \{\sup_{s \in [h,\dred{\tilde{s}}+h]} X(s) \geq u,\sup_{[0,\Ss -h'] \setminus [h,\dred{\tilde{s}}+h]} X(s) \leq u\}$.
Similarly, we have
\begin{align*}
\mu(h) 
& = \Pr\{A(h,h')\} - \Pr\{A(h,h'), \sup_{s \in [\Ss - h',\Ss]} X(s) \geq u\}.
\end{align*}

\begin{figure}[H]
\begin{center}
\def\h{1}
\def\stilde{3}
\def\hprime{1.6}
\def\CapS{10}
\def\step{0.2}
\begin{tikzpicture}
    \draw ({-\hprime - \step},0) --({\CapS + \step},0);
    \node at ({-\hprime},0) {$|$};
    \node[below = 6pt] at ({-\hprime},0) {\footnotesize{$-h'$}};
    \node at ({0},0) {$|$};
    \node[below = 6pt] at (0,0) {\footnotesize{$0$}};
    \node at ({\h},0) {$|$};
    \node[below = 6pt] at ({\h},0) {\footnotesize{$h$}};
    \node at ({\h + \stilde},0) {$|$};
    \node at ({\CapS - \hprime},0) {$|$};
    \node[below = 6pt] at ({\CapS - \hprime},0) {\footnotesize{$\Ss - h'$}};
    \node at ({\CapS},0) {$|$};
    \node[below = 6pt] at ({\CapS},0) {\footnotesize{$\Ss$}};
    \node at ({\h},0) {$|$};    \node[below = 6pt] at ({\h + \stilde},0) {\footnotesize{$\tilde{s}+h$}};
    \foreach \xp in {0.2,0.4,...,0.8}{\node[color = red] at (\xp,0) {/};}
    \foreach \xp in {1.2,1.4,...,3.8}{\node[color = blue] at (\xp,0) {/};}
    \foreach \xp in {4.2,4.4,...,8.2}{\node[color = red] at (\xp,0) {/};}
    \foreach \xp in {8.6, 8.8,...,9.8}{\node[color = orange] at (\xp,0) {/};}
    \foreach \xp in {-1.4,-1.2,...,-0.2}{\node[color = black] at (\xp,0) {/};}
\end{tikzpicture}
\end{center}
\end{figure}

The difference $\mu(h + h') - \mu(h)$ is thus
\begin{align*}
   \mu(h + h') - \mu(h) 
   & = \Pr\{\sup_{[0,\Ss -h'] \setminus [h,\dred{\tilde{s}}+h]} X(s) \leq u, \sup_{s \in [h,\dred{\tilde{s}}+h]} X(s) \geq u,\sup_{s \in [\Ss - h',\Ss]} X(s) \geq u \} \\
   & - \Pr\{\sup_{[0,\Ss -h'] \setminus [h,\dred{\tilde{s}}+h]} X(s) \leq u, \sup_{s \in [h,\dred{\tilde{s}}+h]} X(s) \geq u,  \sup_{s \in [-h',0]} X(s) \geq u\}
\end{align*}
The condition $\sup_{[0,\Ss -h'] \setminus [h,\dred{\tilde{s}}+h]} X(s) \leq u$ defines a hitting scenario $H_1$ for which $X$ exceeds $u$ exclusively on $[h,\dred{\tilde{s}}+h]$ and $[-h',0]$. 
Following Assumption \ref{assump:pi_decreasing}, we have
\begin{equation}
\label{equ:inequalit proof them}
\Pr\{\sup_{s \in [h,\dred{\tilde{s}}+h]} X(s) \geq u, \sup_{s \in [\Ss - h',\Ss]} X(s) \geq u, H_1\} \geq \Pr\{\sup_{s \in [h,\dred{\tilde{s}}+h]} X(s) \geq u, \sup_{s \in [- h',0]} X(s) \geq u, H_1\}
\end{equation}
if
$$
\text{dist}([h,\dred{\tilde{s}}+h],[-h',0]) = h  > \text{dist}([h,\dred{\tilde{s}}+h],[\Ss-h',\Ss]) = S - h' - (\dred{\tilde{s}}+h),$$
or equivalently if $ h > (\Ss - h' - \tilde{s})/2$. Inequality \eqref{equ:inequalit proof them} is reversed when $ h < (\Ss - h' - \tilde{s})/2$, which concludes the proof of Theorem \ref{th: de novo}.

\section{Qq-plots for the station tail distributions}\label{app: qqplots}
\begin{figure}
\begin{center}
\begin{tabular}{cc}
\includegraphics[width=0.4\textwidth]{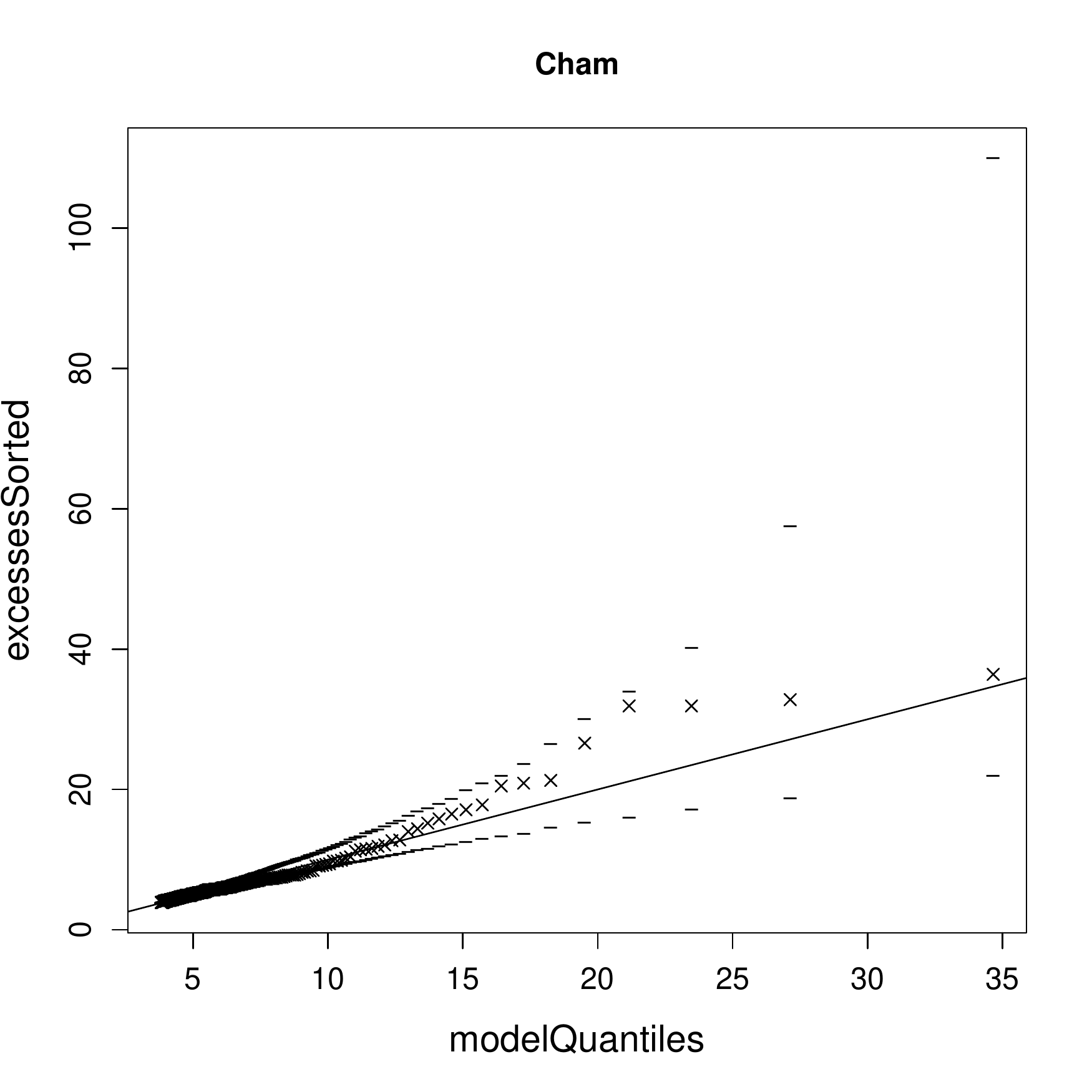} & \includegraphics[width=0.4\textwidth]{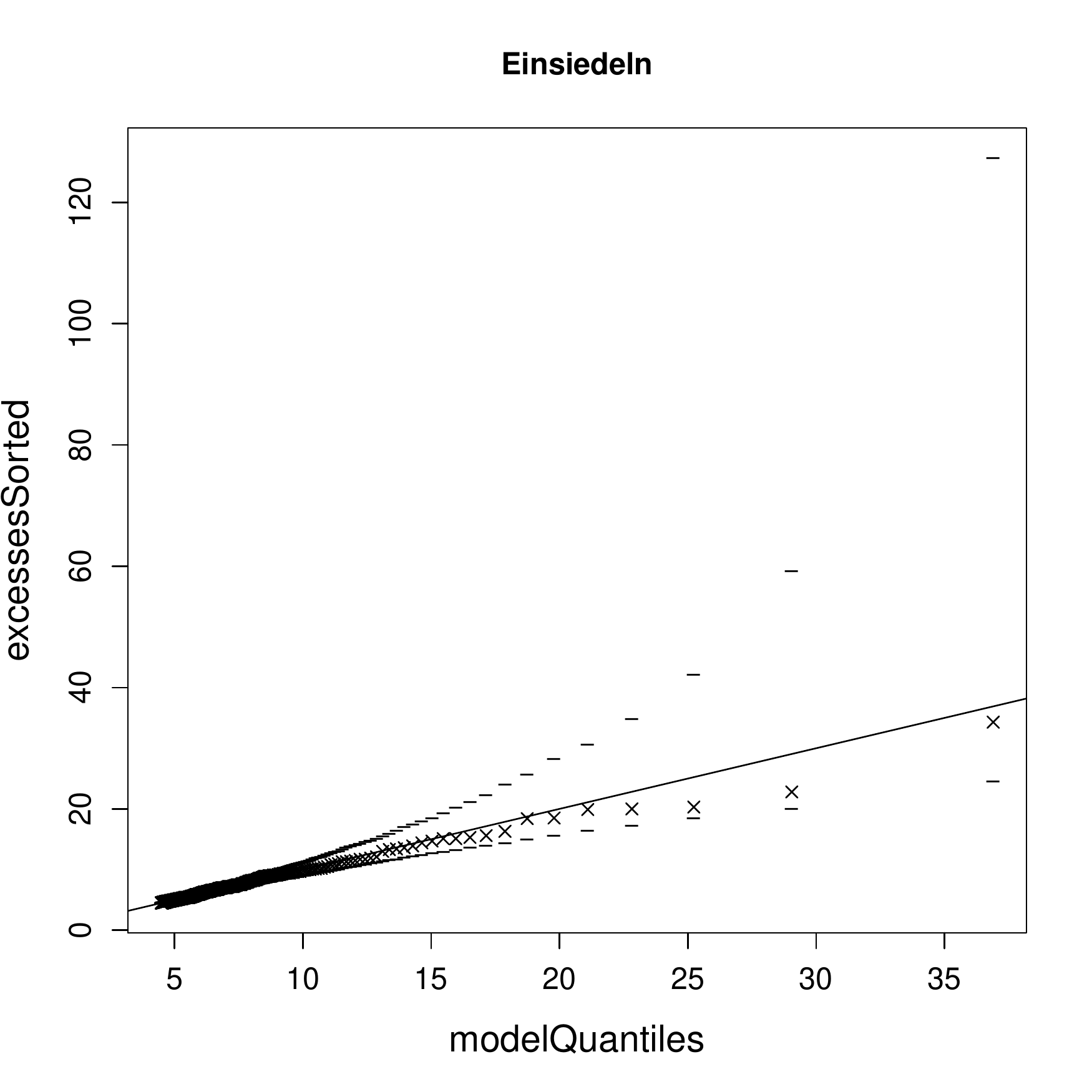}\\ 
\includegraphics[width=0.4\textwidth]{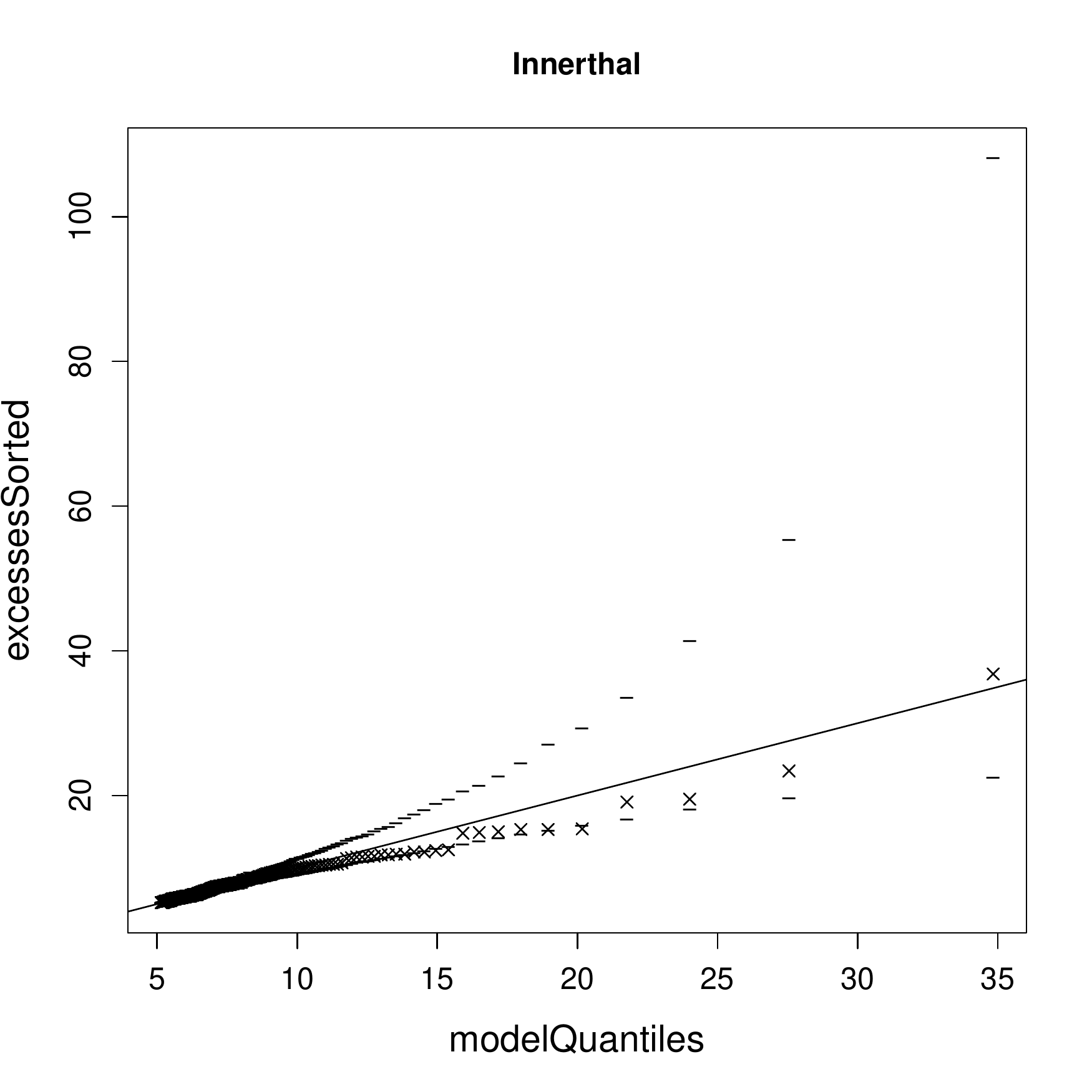} & \includegraphics[width=0.4\textwidth]{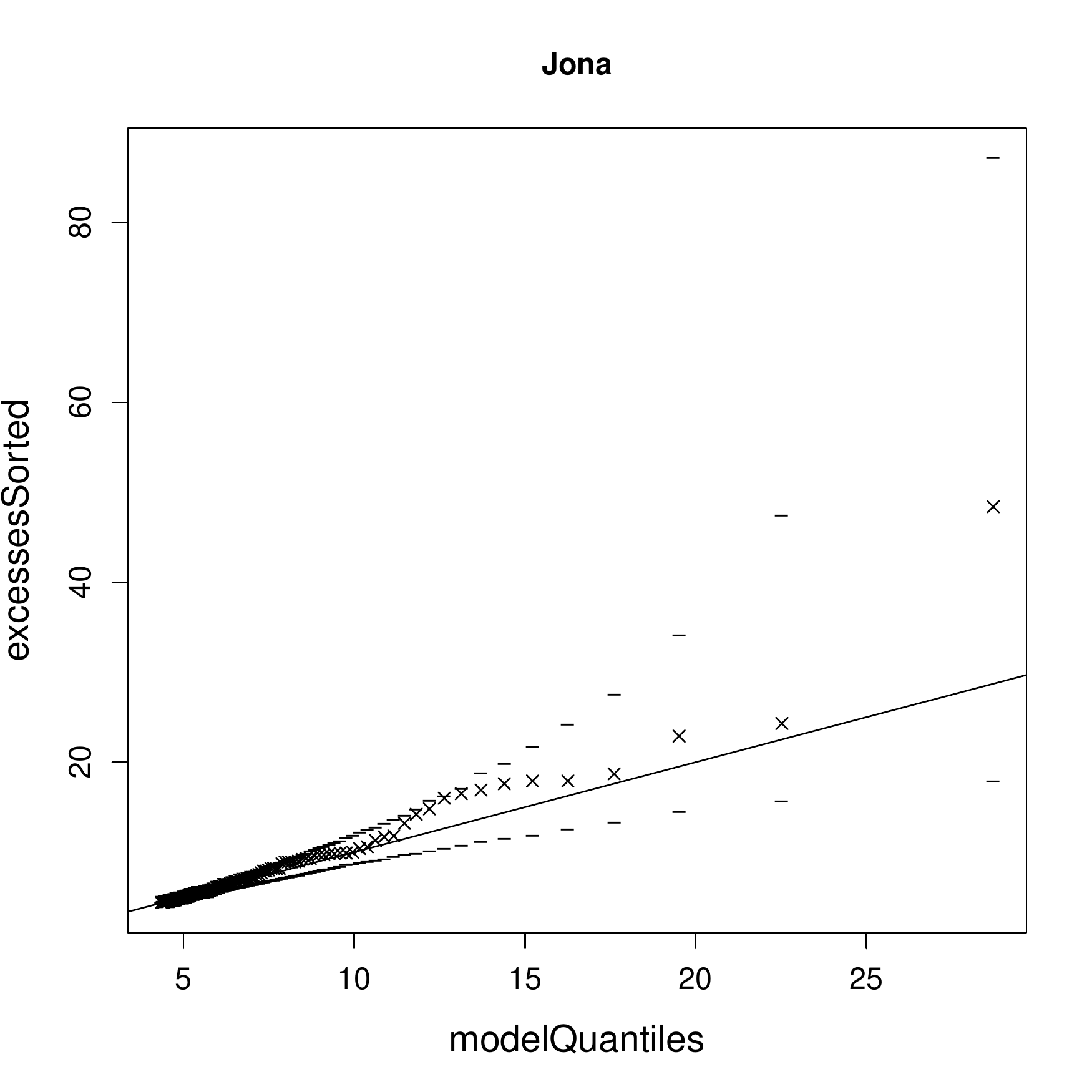}\\ 
\includegraphics[width=0.4\textwidth]{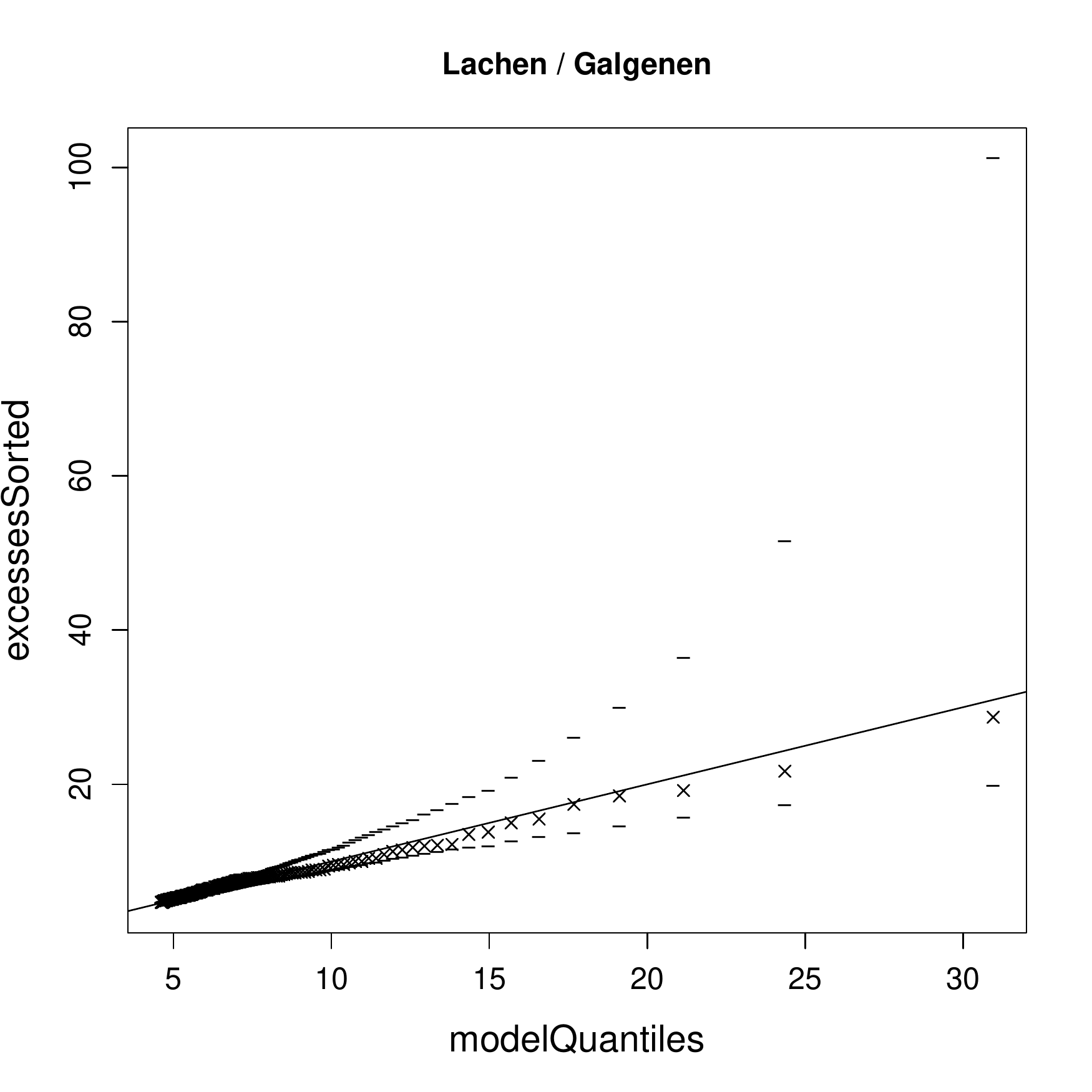} & \includegraphics[width=0.4\textwidth]{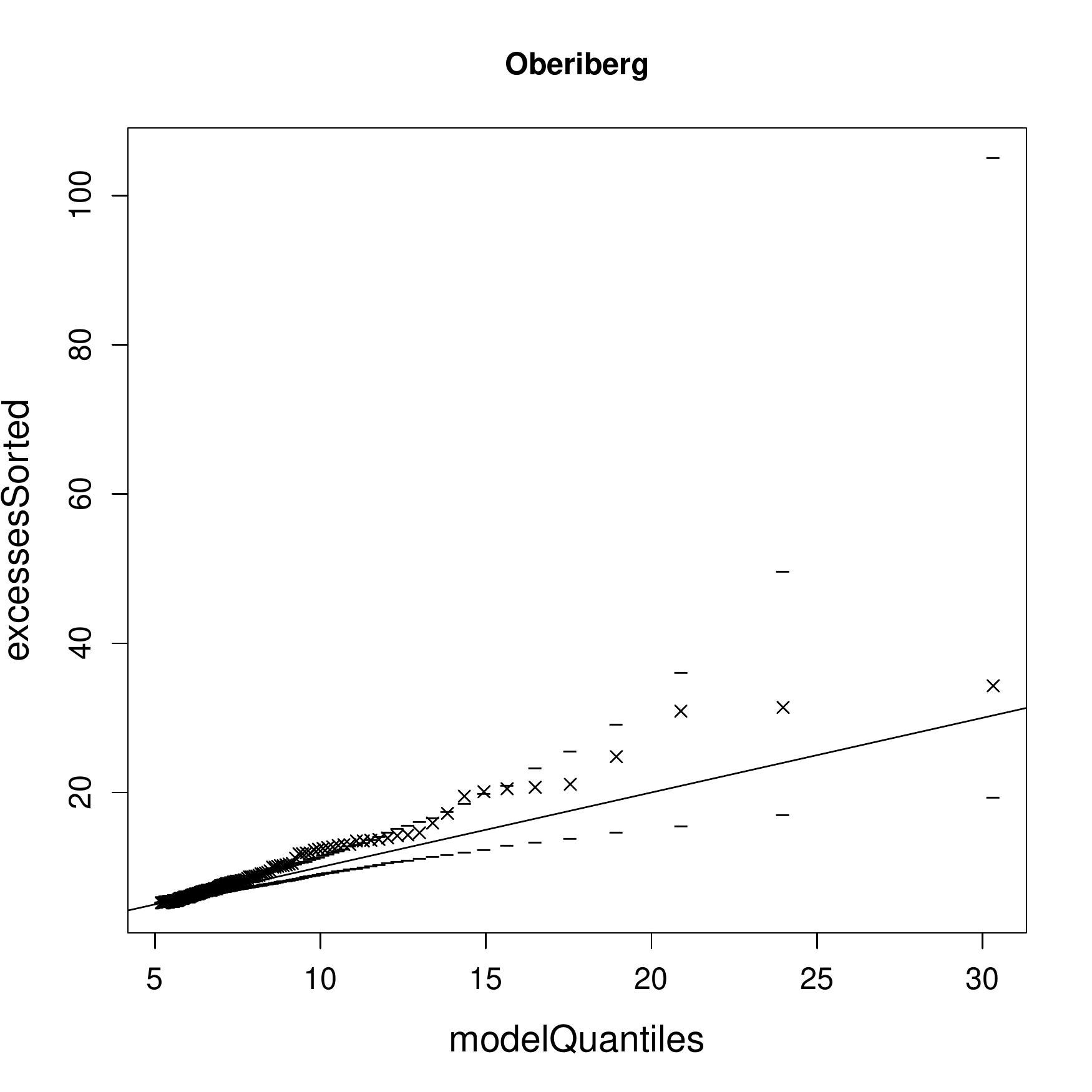}\\ 
\end{tabular}
\end{center}
\caption{QQ-plot of the fitted model for the tail distribution of the $6$ first, out of $14$, stations used in the analysis with $95\%$ confidence intervals obtained by parametric bootstrap.}
\end{figure}

\begin{figure}
\begin{center}
\begin{tabular}{cc}
\includegraphics[width=0.4\textwidth]{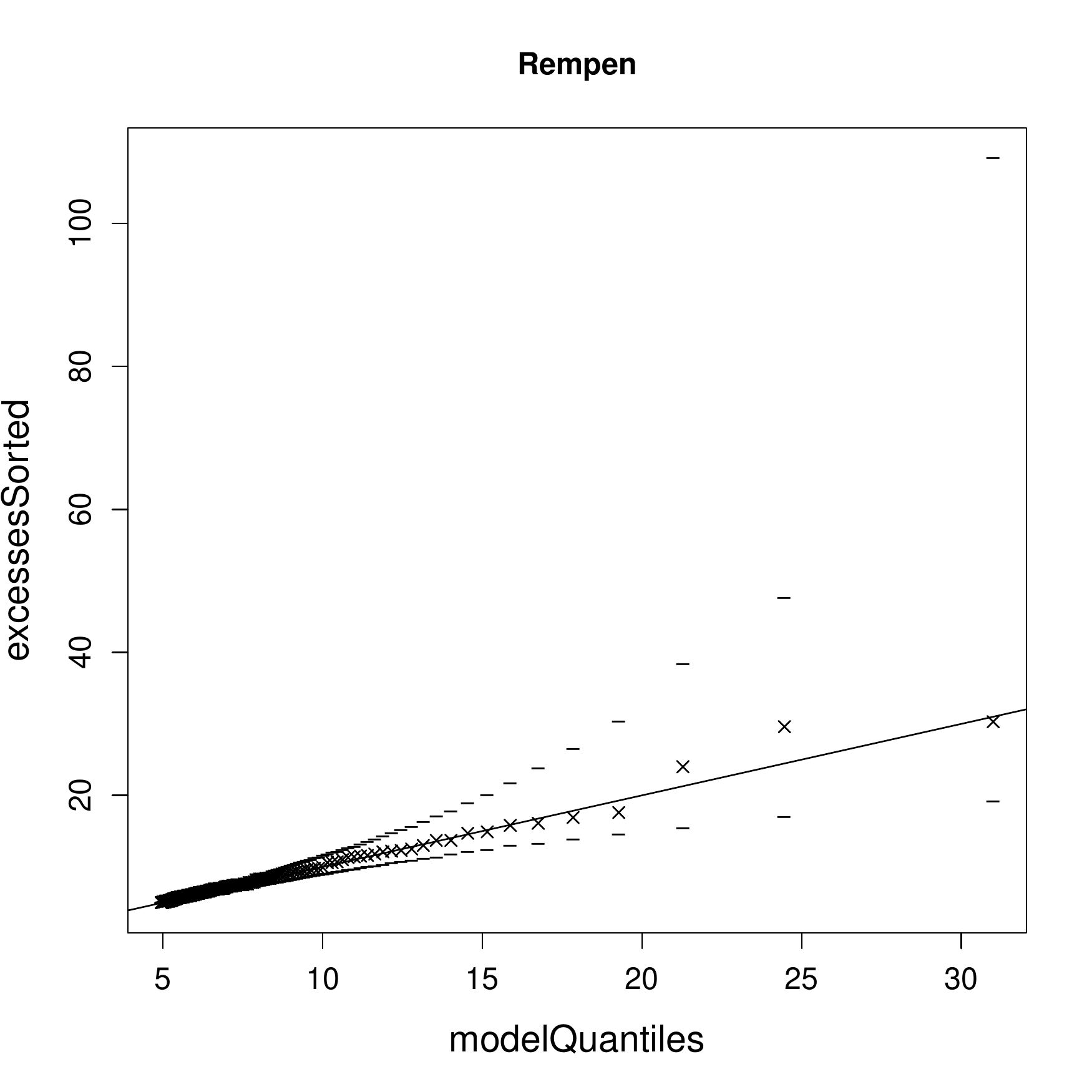} & \includegraphics[width=0.4\textwidth]{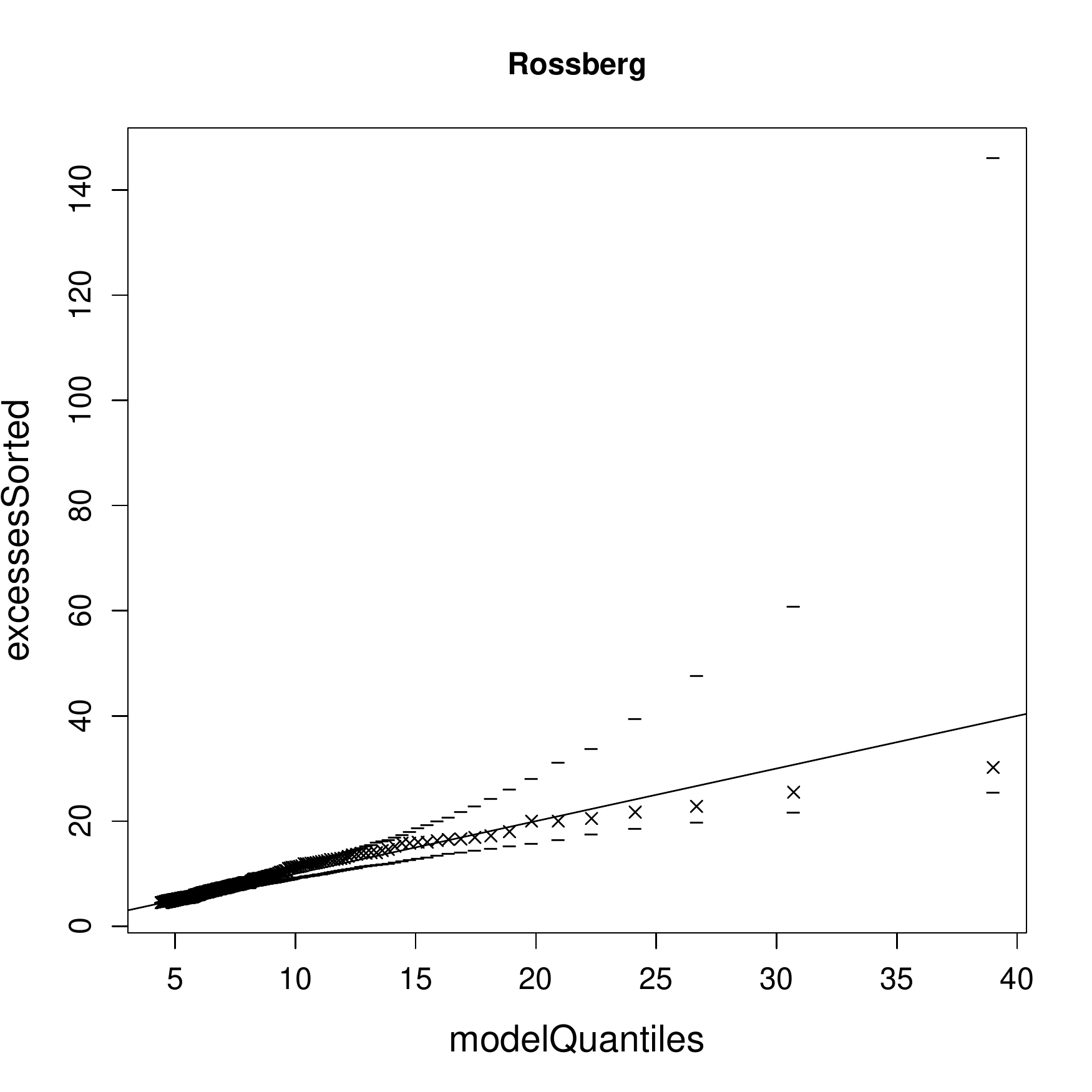}\\ 
\includegraphics[width=0.4\textwidth]{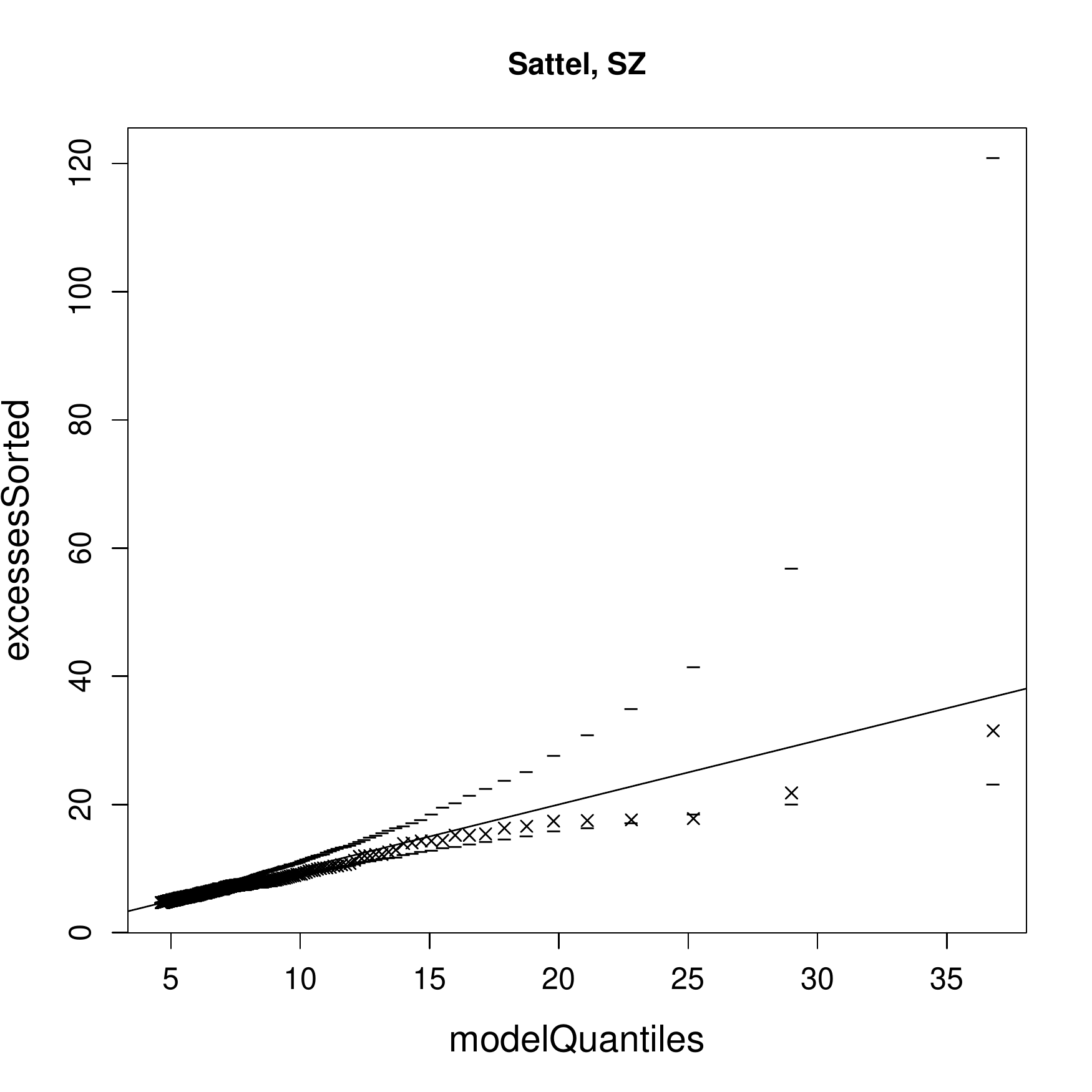} & \includegraphics[width=0.4\textwidth]{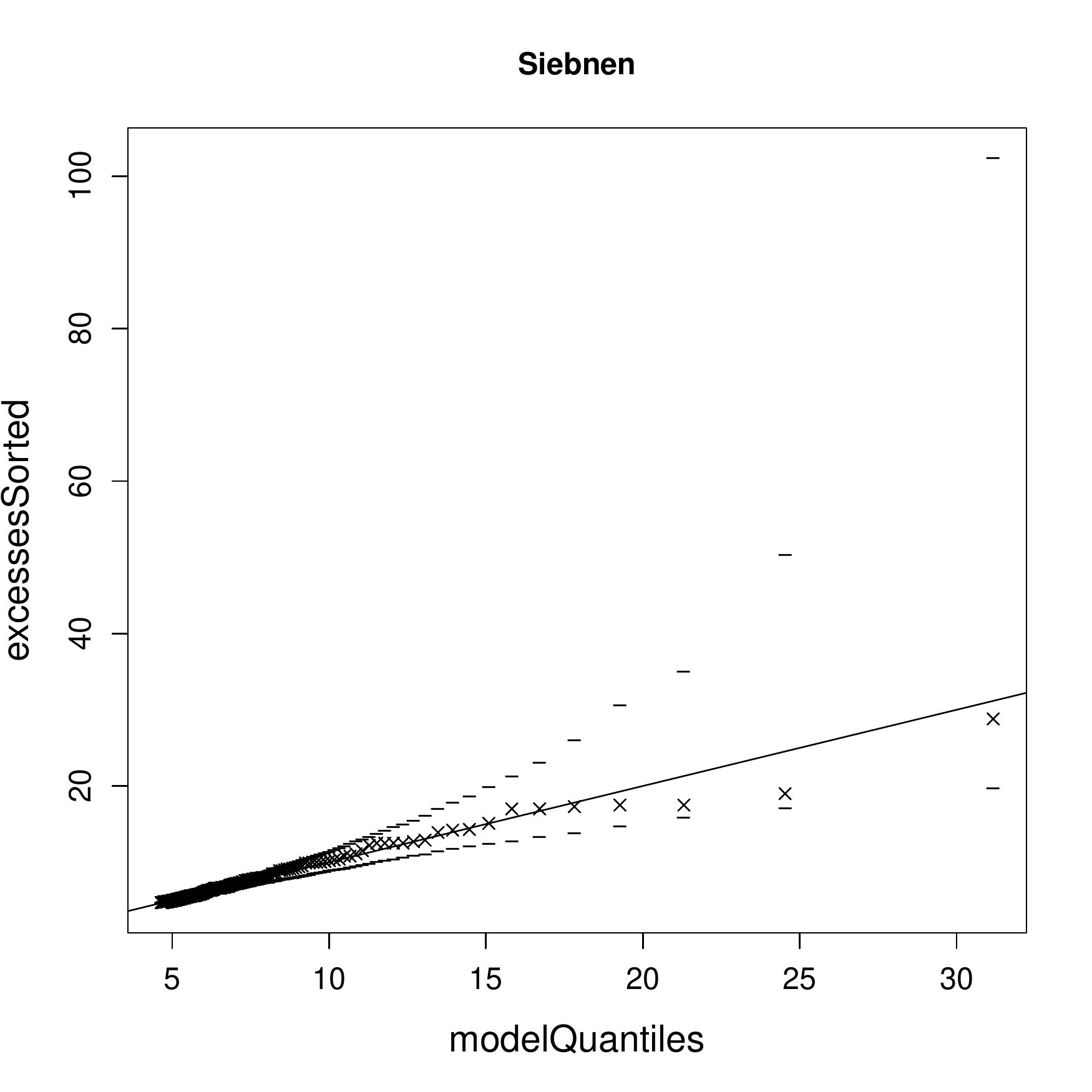}\\ 
\includegraphics[width=0.4\textwidth]{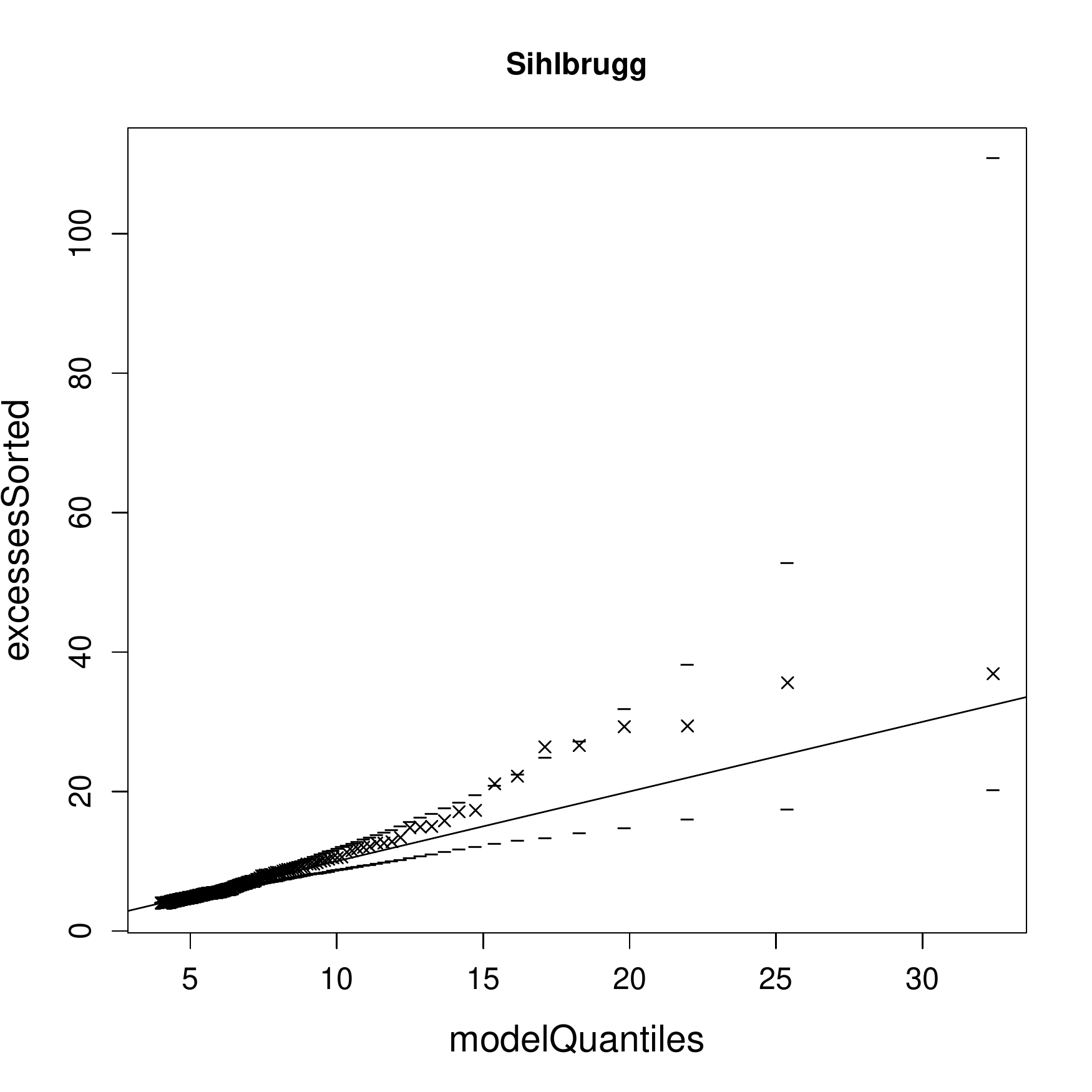} & \includegraphics[width=0.4\textwidth]{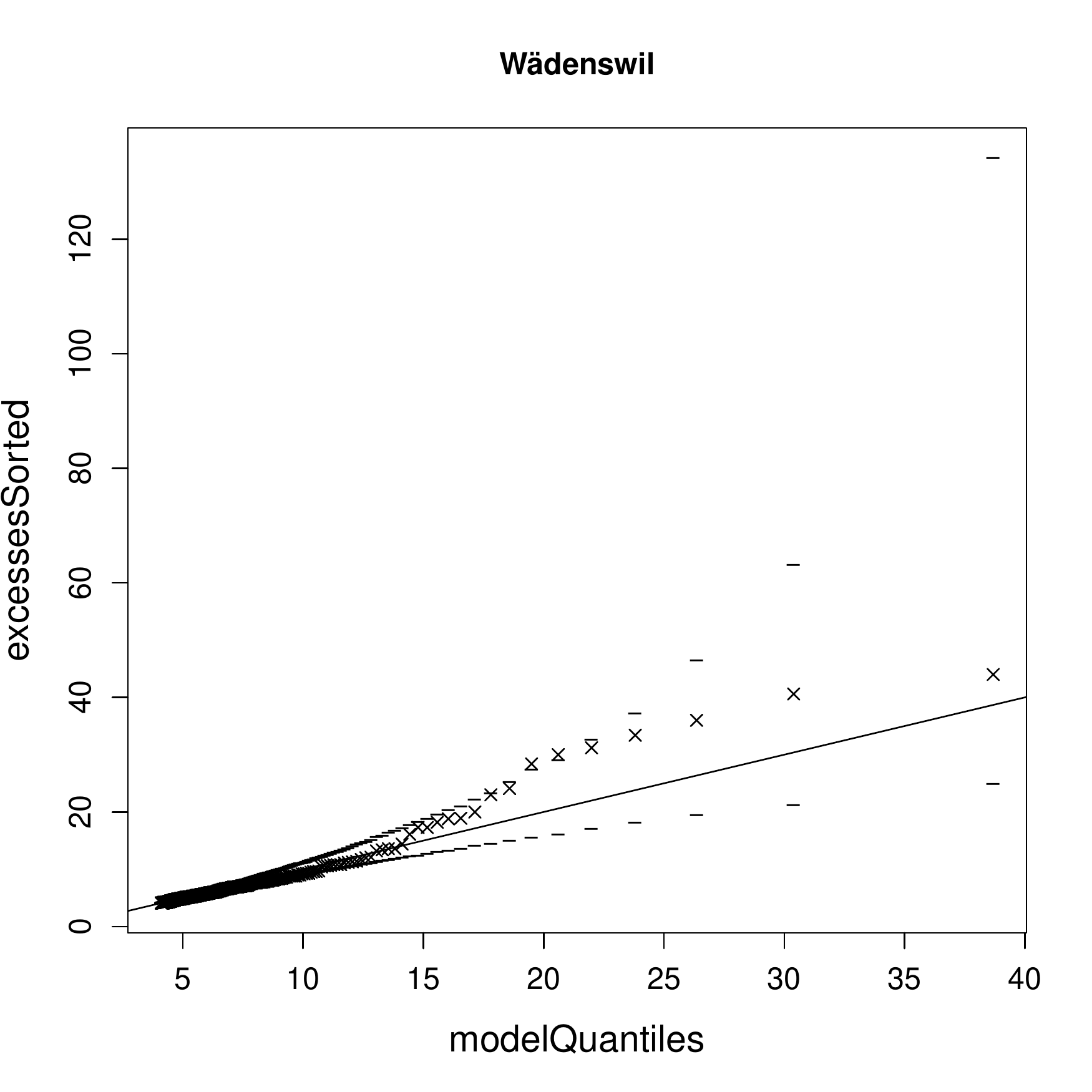}\\ 
\end{tabular}
\end{center}
\caption{QQ-plot of the fitted model for the tail distribution of stations 7 to 12 used in the analysis with $95\%$ confidence intervals obtained by parametric bootstrap.}
\end{figure}

\begin{figure}
\begin{center}
\begin{tabular}{cc}
\includegraphics[width=0.4\textwidth]{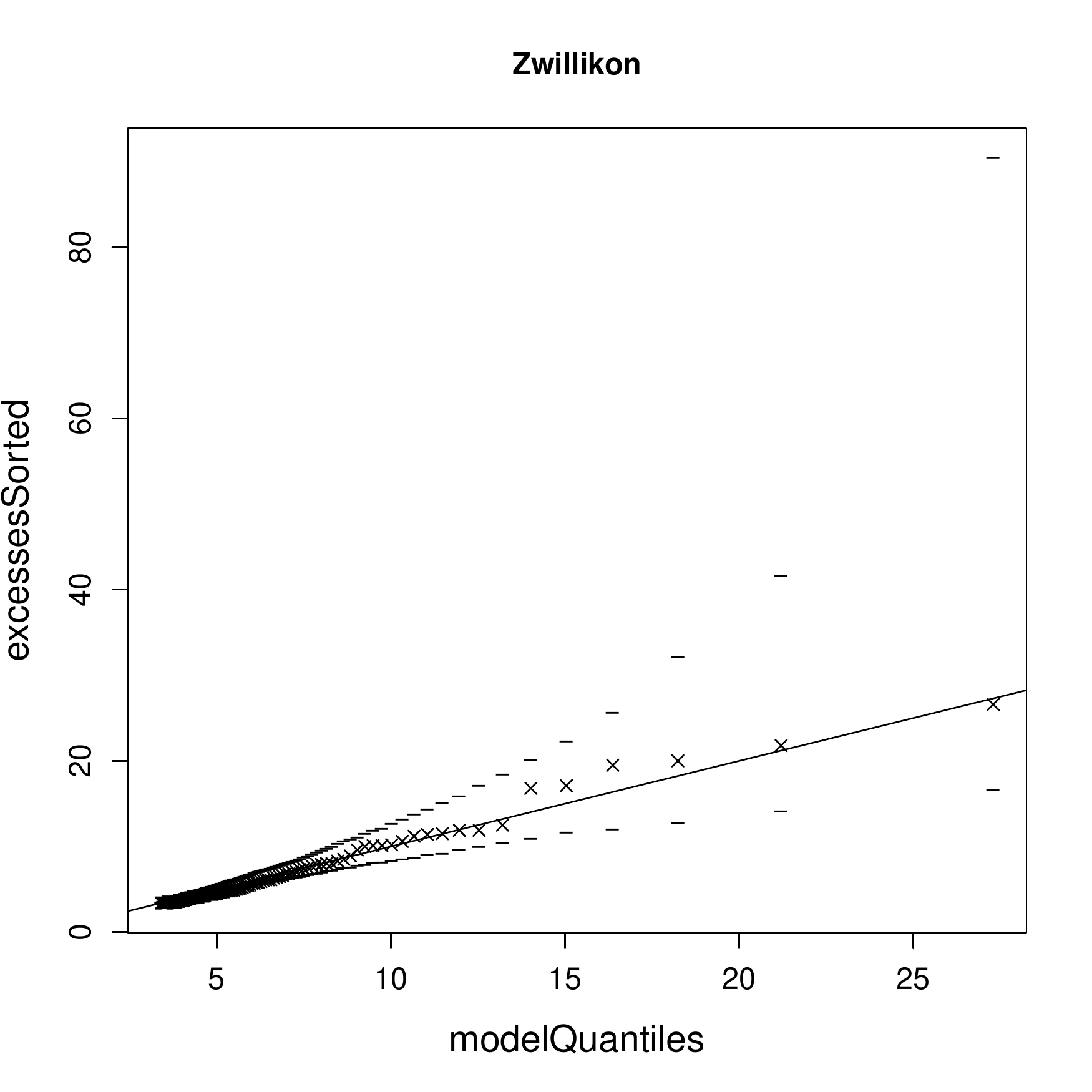}&
\includegraphics[width=0.4\textwidth]{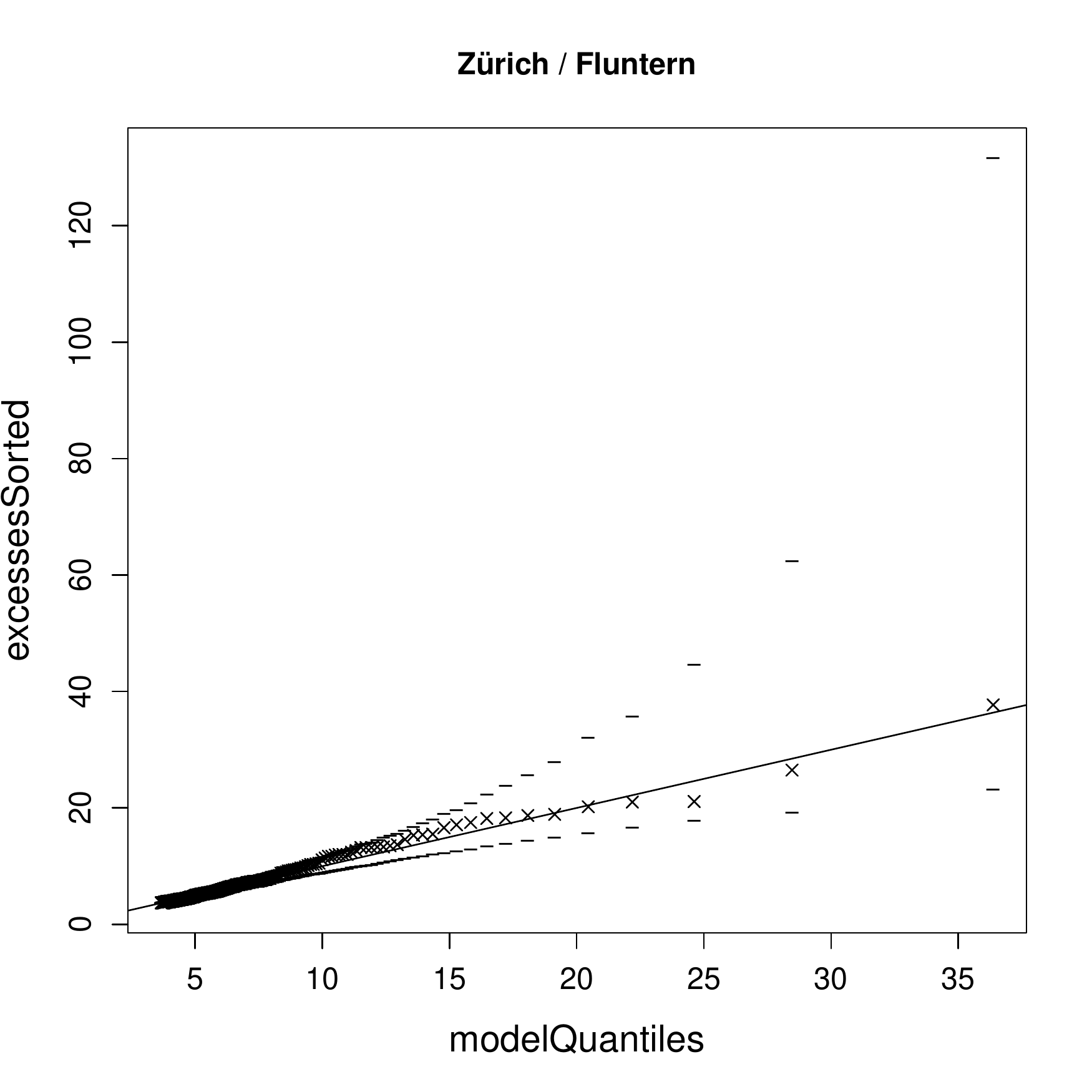}
\end{tabular}
\end{center}
\caption{QQ-plot of the fitted model for the tail distribution of the last 2 stations used in the analysis with $95\%$ confidence intervals obtained by parametric bootstrap.}
\end{figure}

\section{Simulation algorithm log-Gaussian generalized $\R$-Pareto processes}
In the following, we use the convention that, for an array $M$, $M[i,]$ and $M[,i]$ denote respectively the $i-$th line and column respectively.
\label{app: algo sim log-Gaussian}
\begin{algorithm}
\SetAlgoLined
Input semi-variogram $\gamma$, grid coordinates $S_{\rm grid}$, risk functional $\R$, and threshold $u$;
Input number of simulations $N$\;
Optional: input parameters $a$, $b$ and $\xi$\;

Compute semi-variogram matrix $\Gamma = \gamma(s_{1:L_{\rm grid}},s_{1:L_{\rm grid}})$\;
Compute conditional covariance matrix $\Sigma_{ij} = \Gamma_{i1} + \Gamma_{j1} - \Gamma_{ij}$\;
Initialise Sims, a matrix of size $N \times L_{\rm grid}$ with all entries equal to $0$\;

Simulate $N$ Gaussian processes with zero mean and covariance $\Sigma$ and set Sims$[,-1] = \text{Gaussian}(0,\Sigma)$\;

  \For{n from $1$ to $N$}{
  Generate unit Pareto variable $R$\;
  Set Sims$[n,] = R \cdot \frac{\exp(\text{Sims}[n,])}{\|\exp(\text{Sims}[n,]) \|_1}$\;
   \While{$\R\left\{({\rm Sims})[n,]^\xi - 1)\frac{a}{\xi} + b\right\} < u$}{
    Generate a new sample from $G' = \text{Gaussian}(0,\Sigma)$\;
    Generate unit Pareto variable $R'$\;
    Set Sims$[n,] = R' \cdot \frac{\exp(c(0,G')}{\|\exp(c(0,G') \|_1}$
   }
   Set Sims$[n,] = (({\rm Sims})[n,]^\xi - 1)\frac{a}{\xi} + b$\;
 }
 
return $S_{\rm samp}$.
 \caption{Simulation of $N$ realizations of log-Gaussian angular measure \citep{DeFondeville2017}.}
 \label{alg: sim log Gaussian}
\end{algorithm}

\end{document}